\newtheorem{theorem}{\bf Theorem}
\newtheorem{proposition}{\bf Proposition}

\let\latexdocument\document
\let\latexarabic\arabic

 \documentclass[a4paper]{article}

\let\document\latexdocument
\let\arabic\latexarabic

\usepackage{amsmath,amsfonts}
\usepackage{mathrsfs}
\usepackage{multirow}
\usepackage{graphics}
\usepackage{geometry}
\usepackage{caption}
\usepackage{subfigure}

\usepackage{tikz}
\usetikzlibrary{patterns}
\usepackage{subfigure}
\usepackage{tkz-euclide}

\usepackage{times}
\usepackage{bm}
\usepackage{natbib}

\usepackage[ruled,noend]{algorithm2e}

\begin{document} 
	
	
	\markboth{L. SUN and G. WEI}{Integer Grid Bridge Sampler}
	\title{An integer grid bridge sampler for the Bayesian inference of 
		incomplete birth-death records}
	
	\author{Lin \uppercase{SUN} \\ Zhongtai Securities Institute for Financial Studies, Shandong University, \\ 
	  	27 Shanda Nanlu, Jinan 250100, P. R. China\\
  	    {\it email: } sunlin0123@hotmail.com\\ \\
    Gang \uppercase{WEI}\footnote{Correspondence author}\\
   School of Mathematics, Shandong University\\
	27 Shanda Nanlu, Jinan 250100, P. R. China\\
{\it email:} gwei@sdu.edu.cn}     
	
%
%
%
%
%
	
	\maketitle

 

\begin{abstract}
	A one-to-one correspondence is established between the bridge path
	space of birth-death processes and  the exclusive union of the product spaces of  simplexes and integer grids. Formulae  are 
	derived for the exact counting
	of the integer grid bridges with fixed number of upward jumps.  Then a uniform sampler over such restricted bridge path space is constructed. This leads to a Monte Carlo scheme, the integer 
	grid bridge sampler,  IGBS, to evaluate the transition probabilities
	of
	birth-death processes. Even the near zero probability of rare event could now be evaluated with
	controlled relative error. The IGBS based Bayesian inference for the incomplete
	birth-death observations is readily performed in demonstrating examples
	and in the analysis of a severely incomplete data set recording a real epidemic event.  
	Comparison is performed with the basic bootstrap filter, an elementary sequential importance resampling algorithm. The haunting filtering failure has found no position in the new scheme. 
	
\end{abstract}


\section{Introduction}

Statistical inference for birth-death processes has been a critical issue of data analysis 
in fields like population dynamics, epidemiology, genomics, queueing theory, physics, and 
mathematical finance etc \citep{novozhilov_biological_2006,allen2010introduction,pfeuffer_extended_2019}. 
The probability laws governing the birth-death processes have been  well studied, and the likelihood function for a complete sample path  admits a simple explicit expression \citep{reynolds_estimating_1973,crawford_computational_2018}. In most practical cases, however, only partial observations of birth-death processes are available. Except for a few models whose birth and death rates are in linear forms, such as the Yule-Furry process \citep{bailey1990elements}, transition probabilities of birth-death processes usually do not
have explicit expressions. As a result, the likelihood inference for such processes
could be arduous. 

For general birth-death processes, \cite{crawford_transition_2012} proposed an algorithm based on the Laplace transform of transition probability functions. The algorithm first constructs expressions of the Laplace transform of
transition probabilities in continued fractions, and then proceeds to  evaluate the inverse transform 
numerically. As a hybrid with the expectation–maximization, EM, algorithm, this scheme was used to find the maximum likelihood estimate of a birth-death process with partial observations   \citep{crawford_estimation_2014}. Furthermore, \cite{ho_direct_2018,ho_birth/birth-death_2018} 
applied this algorithm successfully in classical epidemiology dynamic models, such as the SIR and SEIR models etc. However, the scheme seems difficult to be applied to multidimensional birth-death processes, for example, the predator-prey model.

Last decades have eye-witnessed the thriving of simulation-based inference methods for birth-death 
processes, especially in contexts where the observations are incomplete or noisy. The most popular technique in this category is the particle filter \citep{kitagawa_non-gaussian_1987,doucet2009tutorial,fearnhead_particle_2018,stocks_iterated_2018}. The popularity of this method lies in its flexibility and convenience in applications. 
Despite its general successes, the so-called filtering failure   \citep{stocks_iterated_2018,stocks_model_2020} is  frequently encountered   in
practice due to weight collapse in the prior-posterior updating procedures. The inapplicability rooted in the mismatch of the proposal
distribution of the importance
sampler embedded in the particle filter for some parameter settings, of which the  posterior samples tend to cluster and induce the damage. 

In the present paper, a  simulation-based method termed as the integer
grid bridge sampler, IGBS, is introduced to evaluate the transition
probabilities of birth-death processes. In a nutshell, the idea is to map the  bridge sample paths with fixed number of upward jumps of birth-death processes to the product space of
a temporal simplex and a spatial integer grid set, and then to conduct the uniform sampling over the simplex and  the
integer grid bridges. This leads to a Monte Carlo scheme to evaluate the  transition probabilities of concern. The algorithm could be extended to the multidimensional situations
without essential difficulties.

In principle the IGBS scheme is a straightforward Monte Carlo technique and thus differs  
essentially from the indirect manner of \citep{crawford_estimation_2014}.  The likelihood inference based upon it was performed successfully for several popular birth-death 
models with partial observations. An attractive feature of the IGBS scheme, in comparison with particle filters, is that the filtering failures no longer haunt.

\section{The Integer Grid Bridge Sampler}
\subsection{Partition of bridge path spaces}
Consider a birth-death process $Y=\{Y_t:\, Y_t\in \mathbb{N}_0,\,t\geq 0\}$, with 0 as 
the minimal index of the states. The transition of $Y$ over the infinitesimal time interval $(t,t+dt]$ is governed by  probability laws:
\begin{equation*}
	\left\{
	\begin{aligned}
		&\text{pr}(Y_{t+dt}=Y_t+1) = \lambda(Y_t;\theta)dt+o(dt),\\
		&\text{pr}(Y_{t+dt}=Y_t-1) = \mu(Y_t;\theta)dt+o(dt),\\
		&\text{pr}(Y_{t+dt}=Y_t) =(1 -  \lambda(Y_t;\theta) - \mu(Y_t;{\theta}))dt+o(dt),\\
	\end{aligned}
	\right.
\end{equation*}
where $\lambda(Y_t;{\theta})\geq 0$ and $\mu(Y_t;{\theta})\geq 0$ denote the birth and death rates respectively, and $\theta$ refers to the parameters. 
In the following $\lambda(Y_t)$ and $\mu(Y_t)$ are frequently used for convenience. 

Though birth-death process paths are piecewise continuous in time, frequently in practice, only incomplete records are available. Let $y_{0:n}=(y_0,\ldots,y_n)^T$ be the 
partially observed sample path of  $Y$, where $y_k=y_{t_k}$ is the value taken by $Y$ at time epoch $t=t_k$ for $k=0,\,\ldots,n$.  Since $Y$ is Markovian, the likelihood function given $y_{0:n}$ takes the form:
\begin{equation}
	L({\theta}\mid y_{0:n}) = \prod_{k=1}^n p(\Delta t_k, y_{k-1}, y_{k})=
	\prod_{k=1}^n p_{ y_{k-1}, y_{k}}(\Delta t_k),    \label{eq_whole_trpdf}
\end{equation}
where $p(t,i,j)=p_{ij}(t)$ represents the transition probability of $Y$ with 
initial state $i$ and terminal state $j$ over time interval $[0,t]$; and $\Delta t_k = t_{k}-t_{k-1}$, $k=0,\,1,\ldots,n$. Thus the evaluation of $L({\theta}\mid y_{0:n})$ is equivalent to  the
calculation of $p_{ij}(t)$. As mentioned previously, apart from a  few cases where $\lambda(Y_t)$ and $\mu(Y_t)$ are linear functions, the explicit expressions of these transition probabilities
cease to exist in theory. Hence numerical schemes have to be constructed for the
likelihood inference.

The simulation based algorithm,  integer grid bridge sampler, IGBS, proposed in the following
is shown 
to be effective in evaluating the transition probabilities of some popular birth-death processes.  The critical idea is outlined as the following three steps.

First, given a complete sample path $\omega=\{\omega_s:\; 0\leq s \leq t\}$ of $Y$ 
bridging states $i$ and $j$, the likelihood  $p(t,i\overset{\omega}{\rightarrow}j)$ has 
explicit expression. Denote the set of all such bridge paths over $[0,t]$ 
by $\Omega_{ij}(t) = \{\omega:\,p(t,i\overset{\omega}{\rightarrow}j)>0\}$. Then  $p_{ij}(t)$ could be expressed as:
\begin{equation}\label{prodj}
	p_{ij}(t) = \int_{\Omega_{ij}(t)} p(t,i\overset{\omega}{\rightarrow}j) d\omega.
\end{equation}

Secondly,   $\Omega_{ij}(t)$ is partitioned according to the numbers of  upward jumps of the sample paths. 

Let $\Omega_{ij}^{B,D}(t)= \{\omega:p(t,i\overset{\omega}{\rightarrow}j)>0, \;\omega \text{ has $B$ upward jumps
	and $D$ downward jumps} \}$. If $j>i$, there are at least $j-i$ upward jumps in  the
bridge path. When  $j\leq i$, the minimal value of $B$ should be $0$. Therefore, $B \in {\cal B}_{ij}=\{(j-i)^+,\ldots,\infty\}$. Since $j-i=B-D$, for a given pair of $(i,j)$,  the value $D=B-(j-i)$ is determined by
$(i,j,B)$. So $\Omega_{ij}^{B,D}(t)$ is written as  $\Omega_{ij}^B(t)$ in the following. Then a mutual
exclusive partition of $\Omega_{ij}(t)$ is in place:
$$\Omega_{ij}(t)=\text{\large $\biguplus_{B\in {\cal B}_{ij}}$}\;\Omega_{ij}^B(t).$$

Thirdly,  a probability measure is endowed over the path space
$ \Omega_{ij}^B(t)$. The distribution is uniform in
the sense that each path in $\Omega_{ij}^B(t)$ has equal chance of being sampled. Denote this uniform distribution by U$(\Omega_{ij}^B(t))$ and  its probability density function
by $h_{ij}^B(t)$.  It will
be shown that $h_{ij}^B(t)$ is constant over $\Omega_{ij}^B(t)$ and formulae are given for its exact value. Then expression (\ref{prodj}) can be further decomposed as:
\begin{equation}
	\label{prodjd}
	p_{ij}(t)  
	= \sum_{B\in {\cal B}_{ij}} \text{ \large $\int_{\Omega_{ij}^B(t)}$} \;p(t,i\overset{\omega}{\rightarrow}j)\; d\omega
	= \sum_{B\in {\cal B}_{ij}}\text{ \large $\int_{\Omega_{ij}^B(t)}$} \left[\frac{p(t,i\overset{\omega}{\rightarrow}j)}{h_{ij}^B(t)}\right] \; 
	h_{ij}^B(t)  \;d\omega.
\end{equation}

Restricted to birth-death processes with no explosions, the probability that $Y$ performs infinitely 
many jumps over a finite time interval is $0$. Therefore a finite subset ${\cal B}^*_{ij}\subset
{\cal B}_{ij} $ is taken for rational approximation, thus  equation
(\ref{prodjd}) now takes the form: 
\begin{eqnarray}
	\label{prodjd2}
	p_{ij}(t) & \approx & \sum_{B\in {\cal B}^*_{ij}}\text{ \large $\int_{\Omega_{ij}^B(t)} $}\left[\frac{p(t,i\overset{\omega}{\rightarrow}j)}{h_{ij}^B(t)} \right]\; 
	h_{ij}^B(t)  \;d\omega .
\end{eqnarray}
Expression (\ref{prodjd2}) implies that a Monte Carlo scheme to evaluate $p_{ij}(t)$ could
be expected. Let $B$ follow the uniform distribution over $ {\cal B}^*_{ij}$.
For each sampled $B$, a bridge path $\omega$ connecting $i$ and $j$ could be generated from U$(\Omega_{ij}^B(t))$. Thus, a Monte Carlo estimator of $p_{ij}(t)$ follows:
\begin{equation}
	\label{igbs}
	\hat{p}_{ij}(t) =\frac{M}{N}  \sum_{k=1}^{N} \frac{p(t,i\overset{\omega^{(k)}}{\longrightarrow}j)}{h_{ij}^{B_k}(t)},
\end{equation}
where $N$ is the sample size, and $M=$ card$({\cal B}^*_{ij} )$, the number of elements in $ {\cal B}^*_{ij} $. The subscript $k$ indicates the $k$th  bridge path sampled from $\Omega_{ij}^{B_k}(t)$. Setting $  {\cal B}^*_{ij} =\{ B\}$, (\ref{igbs}) reads out
the estimate of $p_{ij}^B(t)$, the probability that $Y$ travels from $i$ to $j$ in
time $t$ with exactly $B$ upward jumps:  
\begin{equation}
	\label{igbs2}
	\hat{p}_{ij}^B(t) =\frac{1}{N}  \sum_{k=1}^{N} \frac{p(t,i\overset{\omega^{(k)}}{\longrightarrow}j)}{h_{ij}^{B}(t)}.
\end{equation}

Two technical details need to be addressed to finalize the simulation algorithm:
\begin{itemize}
	\item [(i)] Given a complete path $\omega$ bridging $i$ and $j$ over time interval $[0,t]$, present the explicit expression of $p(t,i\overset{\omega}{\rightarrow}j)$;
	\item [(ii)] Define the uniform  distribution U$(\Omega_{ij}^B(t))$ over the bridge path
	space, 
	evaluate $h_{ij}^B(t)$, and present a sampler for this bridge path simulation. 
\end{itemize}

These issues are settled in the following consecutive subsections.

\subsection{Likelihood of a complete sample path}
The birth-death processes have the fundamental feature of local  spatial-temporal  conditional independence \citep{norris_markov_1998,allen2010introduction}. Given the 
current state $Y_t$, the waiting time $W=T-t$ before next jump (at time $T$) obeys an exponential distribution with rate depending on $Y_t$. When the next jump occurs, the direction is governed by the embedded Markov chain, whose transition matrix is formed by
simple manipulation of birth-death rate functions.  These two random  variables,  the waiting time and the jump direction of next event,  are independent conditional upon $Y_t$. 
Systematic treatment of this issue could be found in 
\citep{crawford_computational_2018}.

More specifically, conditioned on $Y_t$,  the waiting time $W$ between $t$ and occurring time of the next jump follows the exponential distribution
\begin{equation}
	\label{W}
	(W\mid Y_t)\sim \text{Exp}(\Lambda_t),\;\;\;\; \Lambda_t=\lambda(Y_t)+\mu(Y_t).
\end{equation} 
Naturally the conditional probability that no jump occurs before $T>t$ is 
\begin{equation}
	\label{W2}
	\exp\left\{-\Lambda_t (T-t) \right\},\;\;\;\; \Lambda_t =
	\lambda(Y_t)+\mu(Y_t).
\end{equation}
Denote the direction of the next jump by $X$, where $X=1$ refers to an upward jump and $X=0$ a downward jump. Given $Y_t$, $X$ is a Bernoulli trial
\begin{equation}
	\label{X}
	(X\mid Y_t) \sim   \text{B}(1,p),\;\;\;\;\;\; p=\frac{\lambda(Y_t)}{\lambda(Y_t)+\mu(Y_t)}
	=\frac{\lambda(Y_t)}{\Lambda_t}.
\end{equation}
Let $\omega$ be the complete path of $Y$ over time interval $[0,t]$ with $\omega_0=i$ and $\omega_t=j$. Suppose that there are $K$ jumps in $\omega$ at jumping times     $\{\tau_k:k=1,\ldots,K,\,0<\tau_1<\tau_2\cdots \tau_K< t\}$. Let
$\tau_{0:K+1}=(\tau_0=0,\tau_1,\ldots,\tau_K,\tau_{K+1}=t)^T$.
Then the values of $\omega$ associated with $\tau_{0:K+1}$ can be written as $\omega_{0:K+1}=(\omega_0=i,\ldots,\omega_K=j,\omega_{K+1}=j)^T$, since $\tau_K$ is the time of the last jump in $\omega$. For the jump directions, 
use $x_k$ to denote the realization of $(X\mid Y_s)$ at $s=\tau_k$, $k=1,2,\ldots, K$.\; Let  $\Delta \tau_k = \tau_k-\tau_{k-1}$, $k=1,2,\ldots,K+1$.   

According to (\ref{W}), (\ref{W2}) and (\ref{X}), the likelihood function of $\omega$ can be expressed as 
\begin{equation}
	\label{cpij_eq2}
	\begin{split}
		p(t,i\overset{\omega}{\rightarrow}j) 
		&= \prod_{k=1}^K \left(\lambda(\omega_{k-1})^{x_k} \mu(\omega_{k-1})^{(1-x_k)}\right)
		\cdot\exp\left\{-\sum_{k=1}^{K+1} (\lambda(\omega_{k-1})+\mu(\omega_{k-1}))\Delta \tau_k\right\}\\
		& = \prod_{k=1}^K \left(\lambda(\omega_{k-1})^{x_k} \mu(\omega_{k-1})^{(1-x_k)}\right)
		\cdot\exp\left\{-\int_{0}^{t} (\lambda(\omega_s)+\mu(\omega_s))ds\right\}\\
		&=\prod_{k=1}^K \left(\lambda(\omega_{k-1})^{x_k} \mu(\omega_{k-1})^{(1-x_k)}\right)\cdot\exp\left\{-S(\omega)\right\},\\
	\end{split}
\end{equation}
where $S(\omega)=\displaystyle{\int_0}^t (\lambda(\omega_s)+\mu(\omega_s))ds
= \sum_{k=1}^{K+1} (\lambda(\omega_{k-1})+\mu(\omega_{k-1}))\Delta \tau_k$.

\subsection{The uniform  distribution  over the bridge path space: U$(\Omega_{ij}^B(t))$}
With the explicit expression of $ p(t,i\overset{\omega}{\rightarrow}j)$ in place (\ref{cpij_eq2}), next issue is to define the uniform distribution
over the bridge path space $\Omega_{ij}^B(t)$, written as U$(\Omega_{ij}^B(t))$, and to construct a related
sampler.

\begin{figure}[htbp]
	\centering
	\subfigure[]{
		\begin{minipage}[b]{0.4\columnwidth}
			\centering
			\includegraphics[height=5cm,width=5cm]{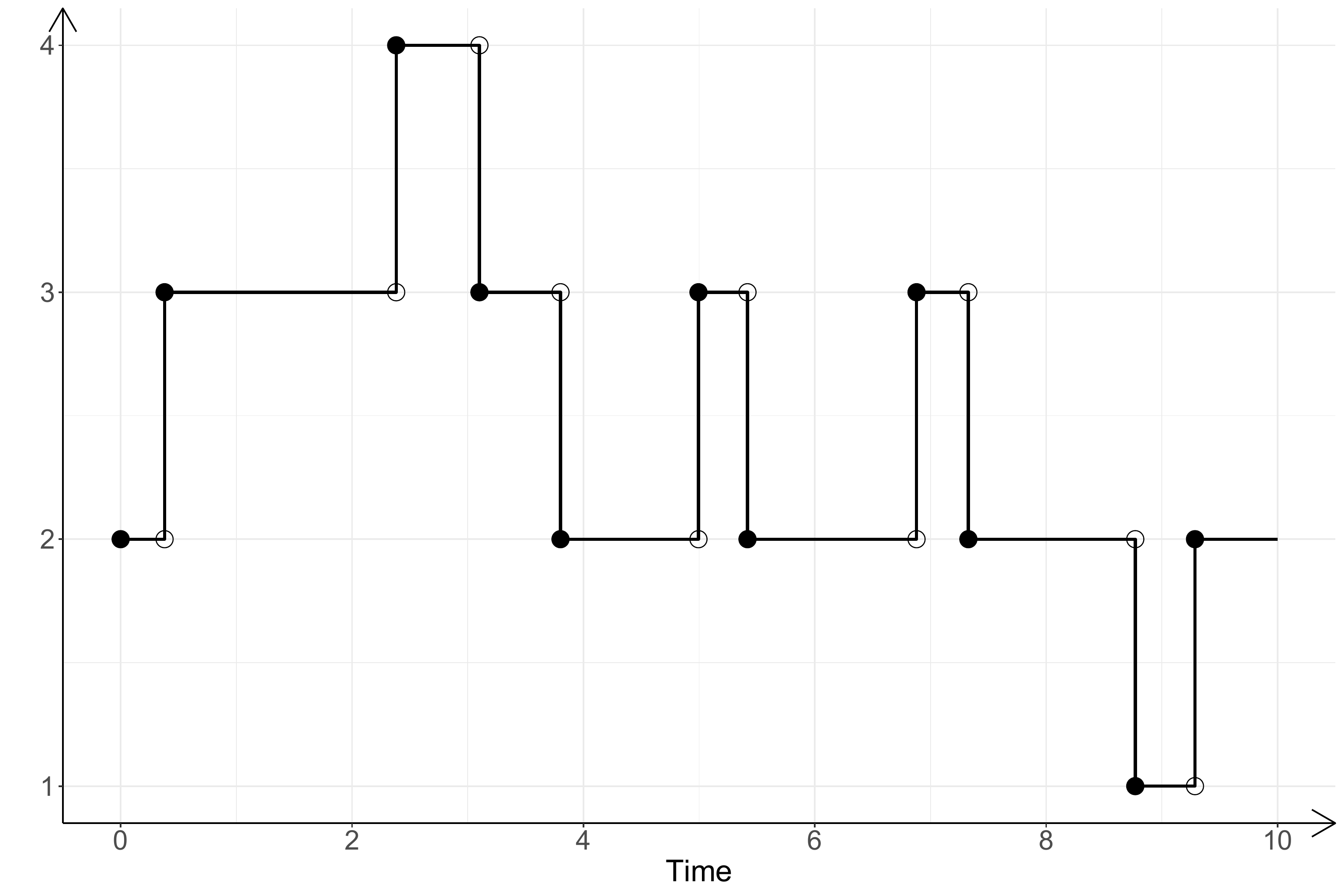}
		\end{minipage}%
	}%
	\hfill
	\subfigure[]{
		\begin{minipage}[b]{0.58\columnwidth}
			\centering
			\includegraphics[height=5cm,width=5cm]{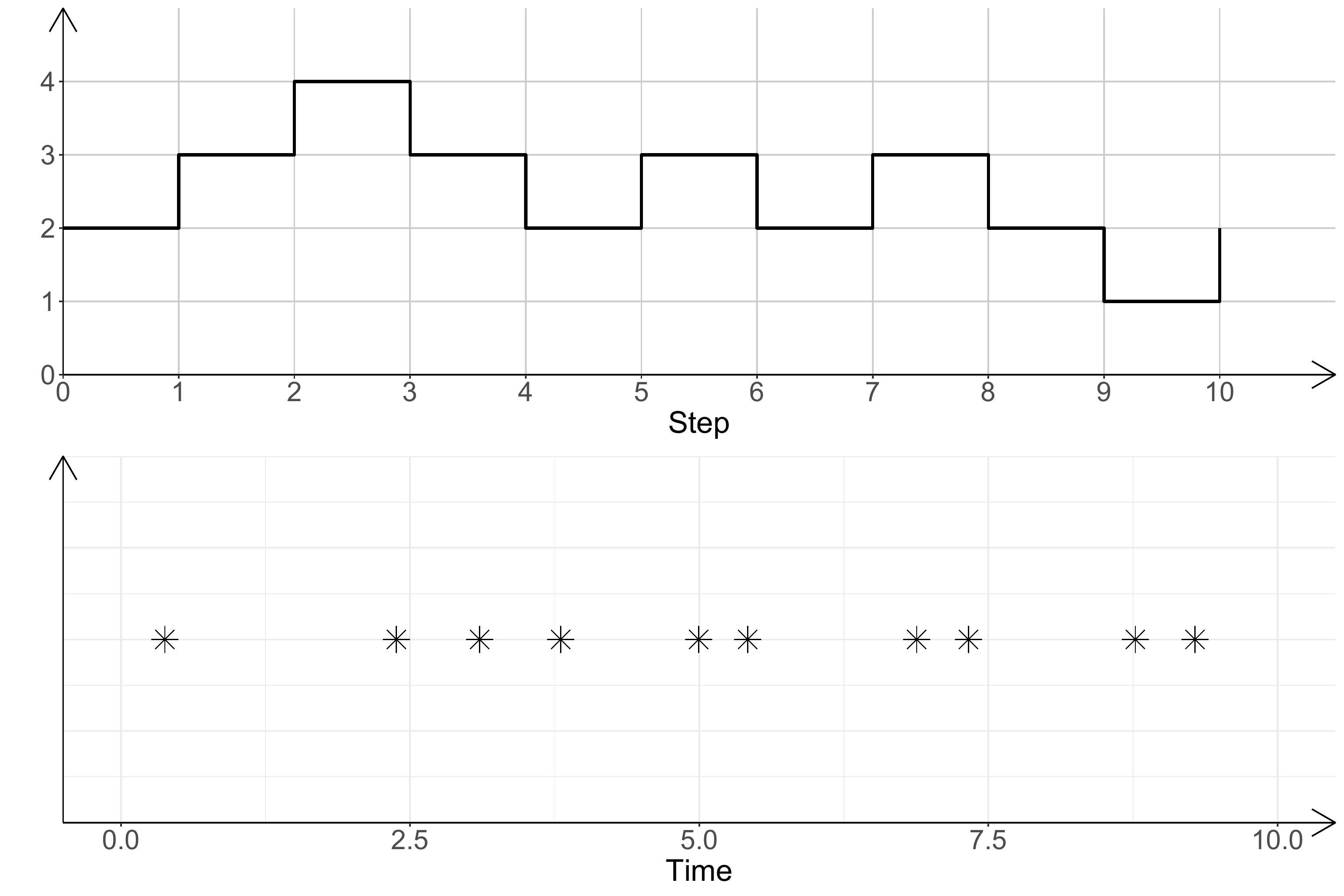}
		\end{minipage}%
	}%
	\centering
	\caption{Illustration of the path decomposition. (a) A typical sample path of birth-death processes. (b) Decomposition of the path in (a), where the upper plot shows the path of the embedded Markov chain and the lower one shows the jumping instants.}
	\label{fig:path_decomp}
\end{figure}

For  a path $\omega$ in $\Omega_{ij}^B(t)$, denote by $\tau_{0:K+1}  =(\tau_0=0,\tau_1,\ldots,\tau_K,\tau_{K+1}=t)^T$ the 
initial and terminal times, and the jump epochs of the  total $K=B+D=2B+i-j$ events. The vector $\omega_{0:K+1} =(\omega_0=i,\omega_1,\ldots,\omega_K=j, \omega_{K+1}=j)^T$ refers to the states  $\omega$ has passed through in the journey from $i$ to $j$, namely the bridge path of the embedded Markov chain over the integer grid spatial-temporal space, i.e. an
integer grid bridge. So, any $\omega$  in $\Omega_{ij}^B(t)$ can be decomposed uniquely into a  2-tuple $(\tau_{0:K+1},\omega_{0:K+1})$. Conversely,   $(\tau_{0:K+1},\omega_{0:K+1})$ can be 
used to reconstruct $\omega$ deterministically. In other words, there is a one-to-one correspondence between $\omega$ and the 2-tuple 
$(\tau_{0:K+1},\omega_{0:K+1})$. This relationship is depicted in Fig. \ref{fig:path_decomp}. \vspace{.3cm}

Let $\mathbb{T}_{ij}^B(t)$ be the set of all possible $\tau_{1:K}$ and $\mathbb{S}_{ij}^B$ be the set of all possible $\omega_{0:K}$. The above discussion implies the 
one-to-one correspondence between $\Omega_{ij}^B(t)$ and  $\mathbb{T}_{ij}^B(t)\otimes \mathbb{S}_{ij}^B$, where $\otimes$ denotes the Cartesian product. Therefore, if
independent samples could be drawn respectively  from $\mathbb{T}_{ij}^B(t)$ and $\mathbb{S}_{ij}^B$ in the classical sense of uniform distributions over a bounded open set in $\mathbb{R}^K$ and a finite
set of integers, then $\omega$ determined by the sampled 2-tuple $(\tau_{0:K+1},\omega_{0:K+1})$ could be taken as a typical sample from U$(\Omega_{ij}^B(t))$.

To endow uniform distributions over $\mathbb{T}_{ij}^B(t)$ and $\mathbb{S}_{ij}^B$, their
different topological features have to be considered separately. For  $\mathbb{T}_{ij}^B(t)$:
\begin{equation}
	\label{tset}
	\mathbb{T}_{ij}^B(t) = \{\tau_{1:K}:0<\tau_1\cdots<\tau_K < t\}.
\end{equation}
This is  an open simplex in $\mathbb{R}^K$, and the uniform distribution over it is well defined with constant probability density function:
\begin{equation} 
	\label{eq_tpdf}
	f_{ij}^B(t) = \frac{K!}{t^K}.
\end{equation}
The simplest sampling scheme for this distribution is to take 
$K$ i.i.d. U$(0,t)$ samples, then use their order statistics as a realization of $\tau_{1:K}$.

As for $\mathbb{S}_{ij}^B$,  it is a finite set, but more details have to be considered 
in response to different birth-death processes, particularly when 
absorbing and reflecting boundaries are of concern. Thus the
lower and upper bounds $(l,u)$ of the integer grid bridge path should be introduced
to discriminate the candidate integer grid bridges. Then it is required  
to count the total number of sample paths and to define the uniform distribution over 
$	\mathbb{S}_{ij}^B(l,u)$. Here $l$ and $u$ are  taken respectively as 
the largest untouchable bound from below and the smallest untouchable bound from above.  When $i$, $j$, and $B$ are given
$(j\leq i+B) $ the   integer grid bridge path would not touch the taboo bounds
$l$ and $u$ if and only if 
$$
u - B > \;\; i \;\; > l + D \;= l + (B-i+j)= \frac12 (l + B +j).  
$$

Uniform samples from $\mathbb{S}_{ij}^B(l,u)$ can be generated from the random shuffling shown below. Represent the jump events as a vector of $\pm 1 $, with  $B$ positive and $D$ negative
elements:
\begin{equation*}
	x = (\underbrace{1,\ldots,1}_{B},\underbrace{-1,\ldots,-1}_{D}).    
\end{equation*}

Take a random  shuffle of $x$ and record the result as $x^\ast=(x_1^\ast,\ldots,x_K^\ast)$. Define $z=(i,x^{\ast})$ and calculate the cumulative summation of its elements. Such operation gives rise a candidate path $\omega_{0:K}=(i,i+x_1^\ast,\ldots,
i+x_1^\ast+\cdots+x_K^\ast=j)^T$. If each element of $\omega_{0:K}$ satisfies the condition that $l<\omega_k <u$, then $\omega_{0:K}\in \mathbb{S}_{ij}^B(l,u)$ and the vector will be kept as a proper sample. Otherwise,  if there is an element in $\omega_{0:K}$ that goes beyond the limits, $\omega_{0:K}$ will be discarded and the random shuffling will 
be repeated  until a satisfying sample bridge is obtained.  

When applying the above procedure, each element of $\mathbb{S}_{ij}^B(l,u)$ shares the same chance of being sampled. The sample should be taken as generated from a uniform distribution and  denoted by $\omega_{0:K}\sim$ U$(\mathbb{S}_{ij}^B(l,u))$. The probability  
density function of U$(\mathbb{S}_{ij}^B(l,u))$ is then the constant equal to the reciprocal of   card$(\mathbb{S}_{ij}^B(l,u))$ evaluated explicitly in the following theorem. 
\vspace{.3cm}
\begin{theorem}
	\label{thm_cs}
	With the notations defined above and let
	\begin{eqnarray*}
		D & = & B+i-j, \;\;\;\;
		K  =  B+D \;\;= 2B+i-j, \\
		B_u  & = &  B-j + u,\;\;\;
		B_l  =  B-j +l, \;\;\;\;
		B_{lu}  =  B - (u-l),\;\;\;\;
		B_{ul}   = B + (u-l).
	\end{eqnarray*}
	The number of bridge paths in $\mathbb{S}_{ij}^B(l,u)$ in
	different situations are give by the formulae:
	\begin{equation}
		\label{C3}
		\text{card}\left(\mathbb{S}_{ij}^B(l,u)\right) =
		\begin{cases}  
			\tbinom{K}{B} , &\text{if}\;\; D+l < i <u-B;\\ \\
			\tbinom{K}{B} -
			\tbinom{K}{B_l}  , &\text{else if}\;\;i\leq\min\left( D+l,u-B-1\right);\\ \\
			\tbinom{K}{B}  -
			\tbinom{K}{B_u}, &\text{else if}\;\;i\geq\max\left( D+l+1,u-B\right);\\ \\ 
			\tbinom{K}{B} -
			\tbinom{K}{B_l}-\tbinom{K}{B_u} \\ 
			\hspace{0.5cm} +  \left[ 	\tbinom{K}{B_{lu}} +
			\tbinom{K}{B_{ul}}   \right]  ,\;\; &\text{else.}\\
		\end{cases}
	\end{equation}
\end{theorem}

\noindent {\bf Proof:}

As shown in the Fig. \ref{fig_mirror}, when there is simply no upper and lower 
bounds, the total number of possible bridges in $\mathbb{S}_{ij}^B(l,u)$,
starting from state $i$ and reaching $j$ in $K$ steps with $B$ 
upward jumps,  
should be the total combination number
$  	\tbinom{K}{B}
$. 
So the claim in the first case of the theorem is obviously true.

\tikzset{global scale/.style={
		scale=#1,
		every node/.append style={scale=#1}
	}
}
\begin{figure}[!ht]
	\centering
	\begin{tikzpicture}[global scale = 0.6]
		\def \r{0.1};
		\draw (0, -7) -- (0, 4);
		\draw (7, -7) -- (7, 4);
		\filldraw (0,0) circle (\r);
		\filldraw (7,1) circle (\r);
		\node at(-0.1, 0)[left] {\Large $i$};
		\node at(7.1, 1)[right] {\Large $j$};
		\draw (0, 0) -- (4,4) -- (7,1) -- (3, -3) -- cycle;
		\draw (0, 2) -- (7,2);
		\node at(7.1, 2)[right] {\Large $u$: \;\;\;\;\;\;\;\; the upper bound};
		\draw (0, -1)--(7,-1);
		\node at(7.1, -1)[right] {\Large $l$:\;\;\;\;\;\;\;\;\;\; the lower bound};
		\filldraw (7,3) circle (\r);
		\node at(7.1, 3)[right] {\Large $j_1=2u-j$};
		\draw[dashed] (0,-4)--(7,-4);
		\node at(7.1, -4)[right] {\Large $2l-u$: \;\; the mirror bound};
		\filldraw (7, -5) circle (\r);
		\node at(7.1, -5)[right] {\Large $j_2=j-2(u-l)$};
		\draw[thick, ->] (0.2,0) parabola (1.9, -0.85);
		\filldraw (2,-1) circle (\r);
		\draw[thick, ->] (2.2, -0.9) parabola (4.4, 1.8);
		\filldraw (4.5, 2) circle (\r);
		\draw[thick, <<-, dotted] (6.8, 3) parabola (4.6, 2.2);
		\draw[thick, <-] (6.8, 1) parabola (4.6, 1.8);
		\draw[thick, ->>>, dashed] (2.2, -1.1) parabola (4.45, -3.8);
		\filldraw (4.5, -4) circle (\r);
		\draw[thick, <<<-, dashed] (6.8, -5) parabola (4.6, -4.2);
		\draw[<-] (0, -6.5) -- (3, -6.5);
		\node at(3.5, -6.5) {\Large $K$};
		\draw[->] (4, -6.5) -- (7, -6.5);
	\end{tikzpicture}
	\caption{Illustration for the reflection principle.}
	\label{fig_mirror}
\end{figure}
\vspace{.3cm} 

When only the upper bound $u$ is taking effect as the second case 
in the theorem, the reflection principle \citep{renault2008lost}
implies that each 
forbidden bridge is uniquely corresponding to a bridge 
in Fig. \ref{fig_mirror} starting
from $i$ and ending in state $j_1=2u-j$ in $K$ steps. Define the 
number of the upward jumps in this mirror bridge as $B_u$, and
$D_u$ for downward ones. Then the equations
\begin{eqnarray*}
	B_u + D_u & = & K = 2B + i -j,\\
	B_u - D_u & = & j_1 - i = 2u-j-i,\\
	\text{sum into}\hspace{2cm}  2 B_u & = &
	2 (B-j+u),\;\;\;\;\text{i.e.}\;\; B_u = B-j+u.
\end{eqnarray*}
This means the total number of forbidden bridges among the candidate paths is $ 	\tbinom{K}{B_u}.  
$
The validity of the claim in the second case of Theorem \ref{thm_cs} 
follows.

The third case of Theorem \ref{thm_cs} could be shown in the same
manner with the upward jump numbers solved as $B_l=B+l-j$.  Then the
number of forbidden bridges among the candidate paths is 
$ 	\tbinom{K}{B_l}.  
$

The last case of Theorem \ref{thm_cs} concerns forbidden bridges
crossing both upper and lower bounds.  Shown in Fig. \ref{fig_mirror} is
the mirror image of a twice flipped forbidden bridge hitting
$l$ first and $u$ later. This implies the double crossing
forbidden bridge has unique correspondence with the bridge
starting from $i$ and ending at $j_2=j-2(u-l)$ in $K$ steps.
Denote the number of upward jumps of the mirror bridge by $B_{lu}$
Then the equations:
\begin{eqnarray*}
	B_{lu} + D_{lu} & = & K = B + D = 2B + i -j,\\
	B_{lu} - D_{lu} & = & j_2 - i = j-i-2(u-l), \\
	\text{ sum into}\hspace{2cm} 
	2 B_{lu} & = & 2 (B-(u-l)),\;\;\;\;\text{i.e.}\;\; B_{lu} = B-(u-l).
\end{eqnarray*}

This means the total number of forbidden bridges is $ 	\tbinom{K}{B_{lu}}.  
$
The situation for the forbidden bridge hitting $u$ first and
$l$ later could be treated in the same way. It is thus
solved with
$B_{ul}=B+(u-l)$ and the number of the related forbidden bridges is
$ 	\tbinom{K}{B_{ul}}. $

Then the claim of the last case in Theorem \ref{thm_cs} follows
by an argument of extracting the  set 
$\mathbb{S}_{ij}^B(l,u)$ from all candidate  bridges do not touch the 
bounds $\{l,u\}$. Thus the proof is completed.

\vspace{.3cm}

So the constant probability density function of U$(\mathbb{S}_{ij}^B(l,u))$
follows: 
\begin{equation}
	\label{eq_spdf}
	g_{ij}^B(l,u) = \frac{1}{\text{card}(\mathbb{S}_{ij}^B(l,u))}.
\end{equation}

Eventually (\ref{eq_tpdf}) and (\ref{eq_spdf}) lead to the constant probability densition function of U$(\Omega_{ij}^B)$:
\begin{equation}
	h_{ij}^B(t) =   f_{ij}^B(t) \; g_{ij}^B(l,u).
\end{equation} 
\vspace{0.05cm}

\subsection{The integer grid bridge sampler, IGBS, and the likelihood inference}

\vspace{0.1cm}
Now the algorithm of integer grid bridge sampler, IGBS,  is in place. Two major 
implementing steps will bear on the mission.  First, generate $N$ i.i.d. samples $\{B_k,\;k=1,\ldots,N\}$  uniformly over 
${\cal B}^*_{ij}$. Secondly, generate bridge paths  $\left\{\omega ^{(k)}\sim  \text{U}(\Omega^{B_k}_{ij}(t)),\;k=1,\ldots,N\right\}$.

With these sample bridges, $p_{ij}(t)$ could be evaluated via (\ref{igbs}). 

\vspace{0.3cm}
\begin{algorithm}[!htp]
	\label{alg:IGBS}
	\caption{IGBS \hspace{.3cm} (integer grid bridge sampler)}
	\label{alg:IGBS}
	\begin{tabbing}
		For {$k=1$ to $N$}\\
		\hspace{.2cm} Draw   $B_k\sim$ U$({\cal B}^*_{ij})$.\\
		\hspace{.2cm} Draw    $\tau_{1:K_k}\sim$ U{\small $( \mathbb{T}_{ij}^{B_k})$}.\\ \hspace{.2cm} Draw   $ \omega_{0:K_k}\sim$ U{\small $(\mathbb{S}_{ij}^{B_k}(l,u))$}.\\
		\hspace{.2cm} Set $\omega^{(k)} = \left[\tau_{0:K_k+1},\;\omega_{0:K_k+1}\right]$
		with $\tau_0=0$, $\omega_{K_k+1} =j$. \\  
		End for $k$\\  
		Evaluate   $ \hat{p}_{ij}(t)$ via equation(\ref{igbs})
	\end{tabbing}
	\vspace*{-10pt}
\end{algorithm}

Particularly if  ${\cal B}^*_{ij}=\{B\}$,  the application of  Algorithm \ref{alg:IGBS}   leads to the evaluation of $p_{ij}^B(t)$.

\section{Application of the IGBS to the Popular Birth-Death Processes}


\subsection{Some classical birth-death process models}

In this subsection, simulation records of several one dimensional birth-death processes are treated with the IGBS method. These processes include the linear population processes and the susceptible-infectious-susceptible, SIS, epidemic model. The specifications of birth and death rate functions for these processes and the corresponding parameter settings adopted for the numerical experiments are tabulated in Table \ref{table:models}.

The linear population process is a birth-death process with immigration,   abbreviated as L-BDI process in the Table \ref{table:models} and here after. Its state space consists of non-negative natural numbers, so proper
upper bounds for population, range of upward jumps of paths $M=$ 
card$({\cal B}^*_{ij})$, and  reflecting and absorbing boundary conditions are introduced for the effective implementation of the IGBS method.
The explicit expressions of the transition probabilities of linear population processes are well known. Hence the applicability of the 
IGBS method could be examined 
straightforwardly as shown in the later part of this subsection. 

\begin{table}[!ht]
	\caption{\it Rate functions and parameter settings of birth-death examples.} 
	\centering
	\begin{tabular}{cccc}
		\hline
		Model & Birth Rate & Death Rate&Parameter settings\\ 
		\hline \\
		L-BDI & $\lambda Y_t +\nu$ &$\mu Y_t$&$(\lambda,\;\mu,\;\nu)=(0.8,\;0.6,\;1.2)$\\ \\
		SIS & $\beta I_t (N_0-I_t)$ & $ \gamma I_t$&$(N_0,\;\beta,\;\gamma)=(30, \;  0.003,\;1)$\\
		\hline
	\end{tabular}
	\label{table:models}
\end{table}

The SIS model, with given finite state space, is one of the simplest but fundamental models in the theory of epidemic dynamics. Consider an isolated population consisting of $N_0$ individuals. The SIS model discriminates the whole population into two compartments. Individuals who are not but may be infected are called the  susceptibles (S),  while those infectious (I) are assumed of 
being able to transmit the epidemic to the susceptible individuals. Numbers of people belonging to the susceptibles and the infectious at time $t$ are denoted by $S_t$ and $I_t$ respectively. Clearly, $S_t+I_t \equiv N_0$. Throughout  the evolution of the epidemic, two kinds of events may happen. One is that a susceptible individual getting infected,  with the chance rate proportional to the product of $S_t$ and $I_t$.
The other is that an infected person getting
recovered and becoming susceptible again, with the chance rate proportional to $I_t$ only. Therefore, the SIS model is essentially  a birth-death process in terms of $I_t$ over the state space $\{0,\ldots, N_0\}$,
with $0$ being absorbing and $N_0$ reflecting. To test the efficiency and
accuracy of the IGBS method for SIS model, the comparison is taken between the estimates obtained through the IGBS method and those given by the large
sample simulations. The probabilities of epidemic termination in different conditions are 
also obtained with aid of the IGBS method.

Some preparatory details for  implementing  the IGBS method  are listed below. 
\begin{itemize}	
	\item L-BDI $(\lambda,\,\mu,\nu)$, $(\lambda\neq \mu)$.
	The transition probability for this birth-death process is a
	binomial mixture of negative binomial distributions:
	\begin{eqnarray*}
		(Y_t\mid Y_0=i) & \stackrel{d}{=} &
		X+Y, \;\;\;\;
		X \sim  \text{B}(i,p),\;\;\;\;
		Y\mid X  \sim  \text{NegBin}(X+\delta,\alpha),\\
		& & \text{where}\;\;\;\; p  =  \frac{\rho (1-c)}{1-\rho c}, \;\;\;\;  \alpha  = \frac{(1-\rho)c}{1-c}, \\
		& & \text{with}\;\; c = \lambda/\mu,  \;\;\;\; \rho  =  \exp(-(\mu-\lambda)t),\;\;\;\; \delta =\nu /\lambda.
	\end{eqnarray*}
	
	The state $0$ is reflecting if $\nu>0$, but absorbing if $\nu=0$.
	\vspace{0.1cm}
	\item SIS. The sample set for $\tau_{1:K}$ shall be $\mathbb{T}_{ij}^B$.  If $0<j\leq N_0$, the sample set for $\omega_{0:K}$ is $\mathbb{S}_{ij}^B(l,u)$, $l=(j-B)^+,\; u=\min(B+i+1,N_0+1) $.  
\end{itemize}

Given the possible range of $B$ and other parameter settings,   Algorithm \ref{alg:IGBS} could now be employed to estimate $p_{ij}(t)$.  
Numerical evaluation is performed for the L-BDI process, with $(t=1,i=5,j=0,\ldots,12)$. Each simulation is performed with sample
size $10^5$.

\begin{figure}[!ht]
	\centering
	
	\subfigure[]{
		\includegraphics[width=0.31\textwidth,height=3.5cm]{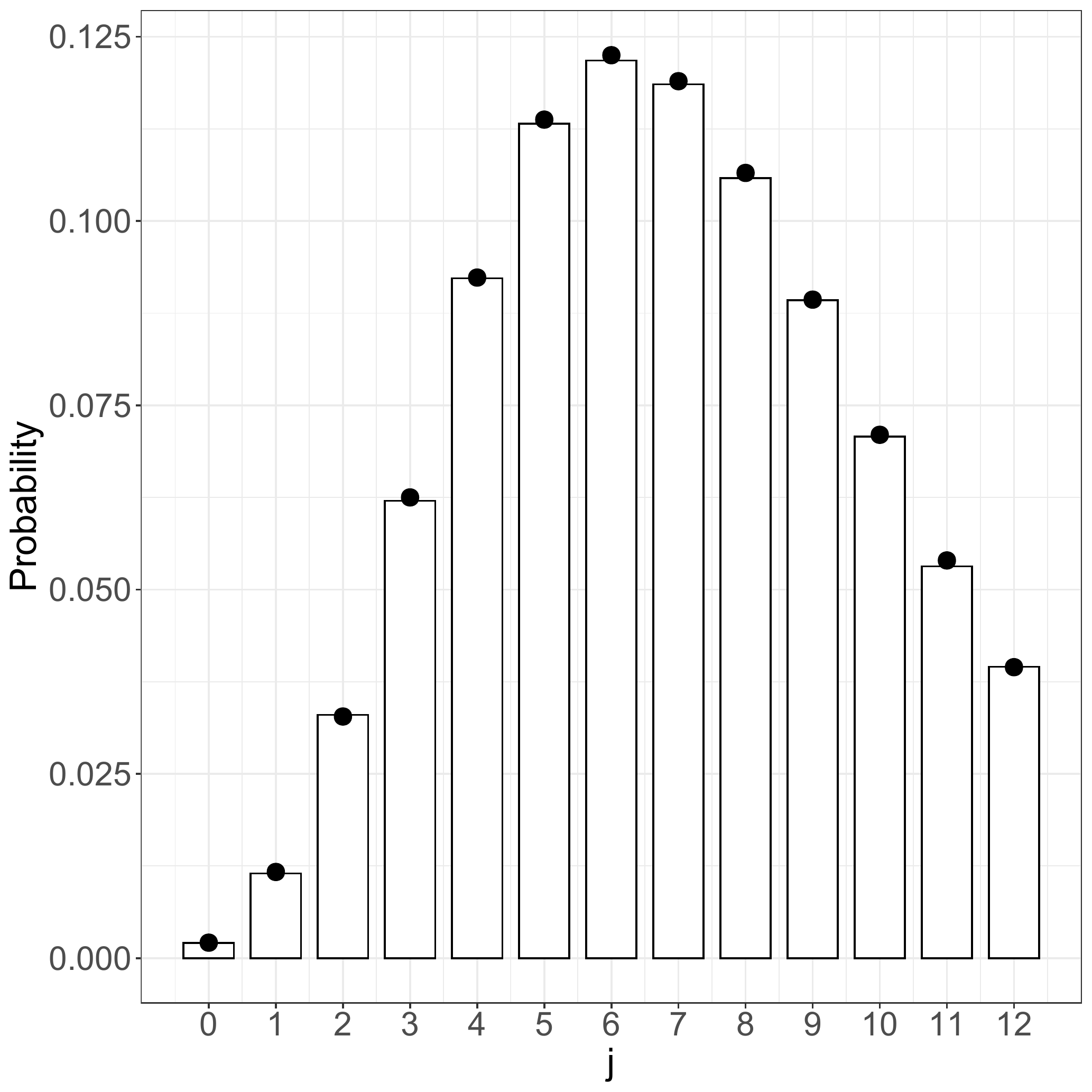}
	} 
	\hfil
	\subfigure[]{
		\includegraphics[width=0.31\textwidth,height=3.5cm]{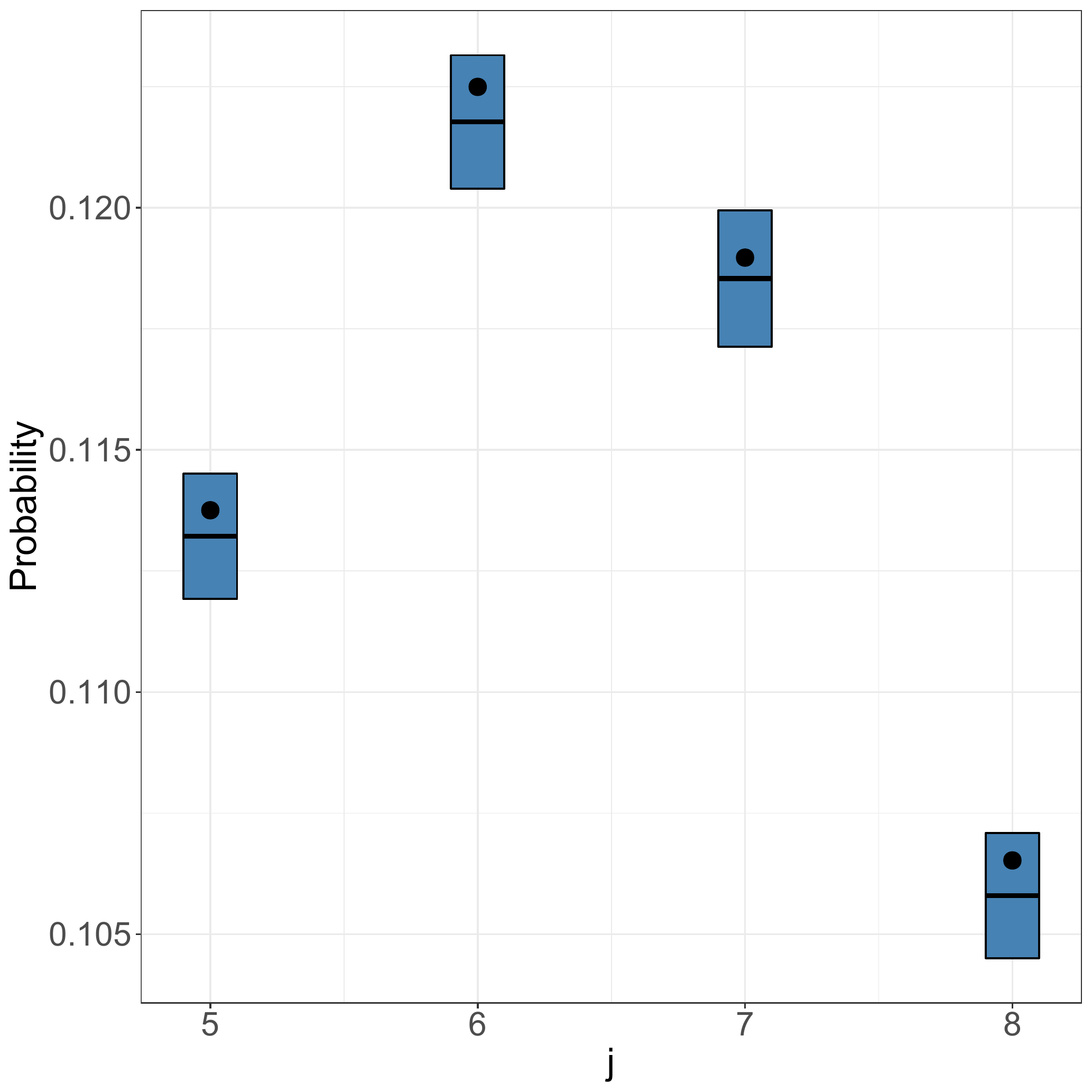}
	}
	
	\caption{
		Transition probabilities of the L-BDI process. The estimates given by the IGBS method are shown by bars and the formula values by black dots. Subfigure (b) is the local zoom of the subfigure (a).  The crossbars indicates the range of $\pm2$ standard deviations of IGBS estimates.}
	\label{fig:num_com1}
\end{figure}
Figure \ref{fig:num_com1} depicts the nice performance of the IGBS method for the linear population processes. 

\begin{figure}[!ht]
	\centering
	\subfigure[]{
		\includegraphics[width=0.31\textwidth,height=4.5cm]{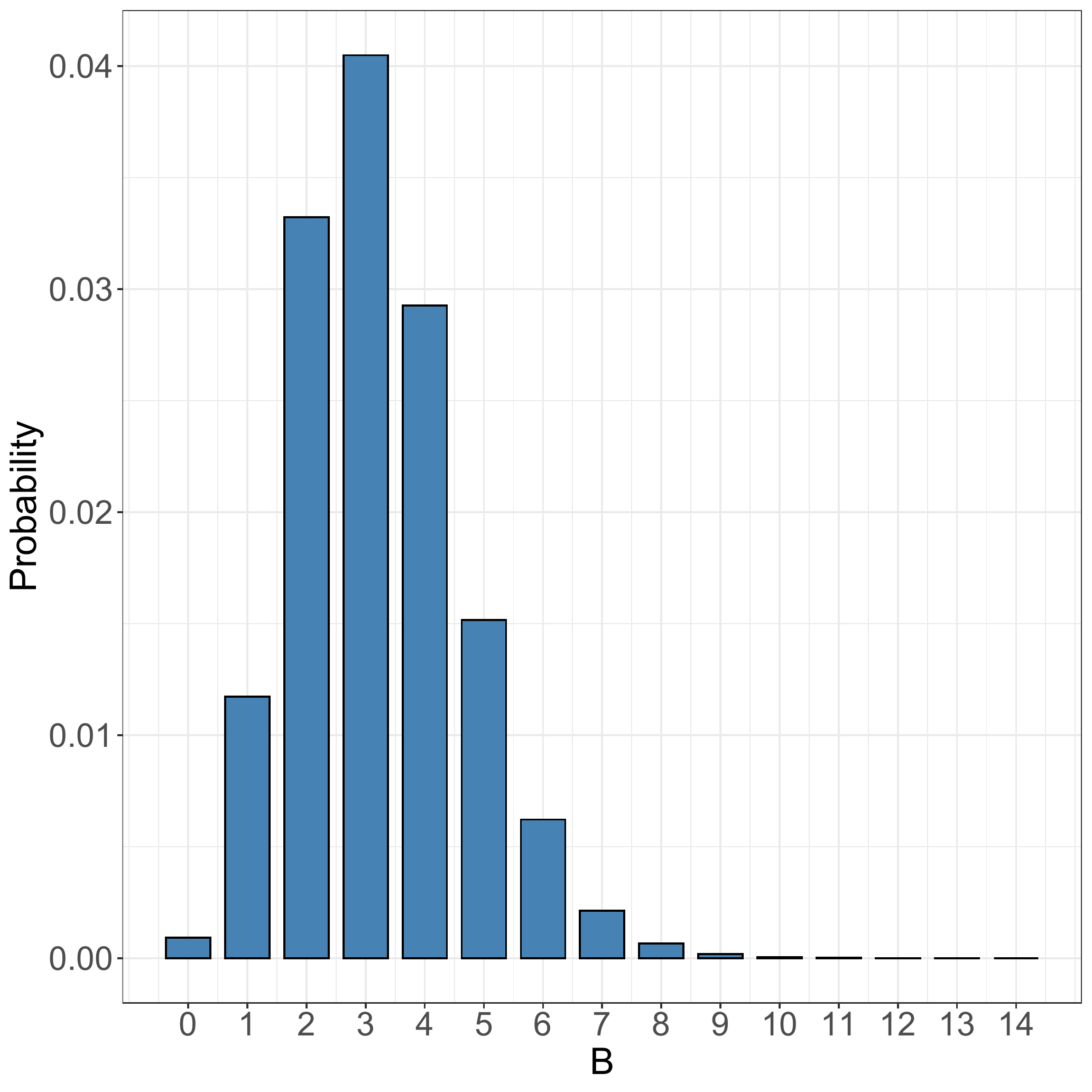}
	}
	\hfil
	\subfigure[]{
		\includegraphics[width=0.31\textwidth,height=4.5cm]{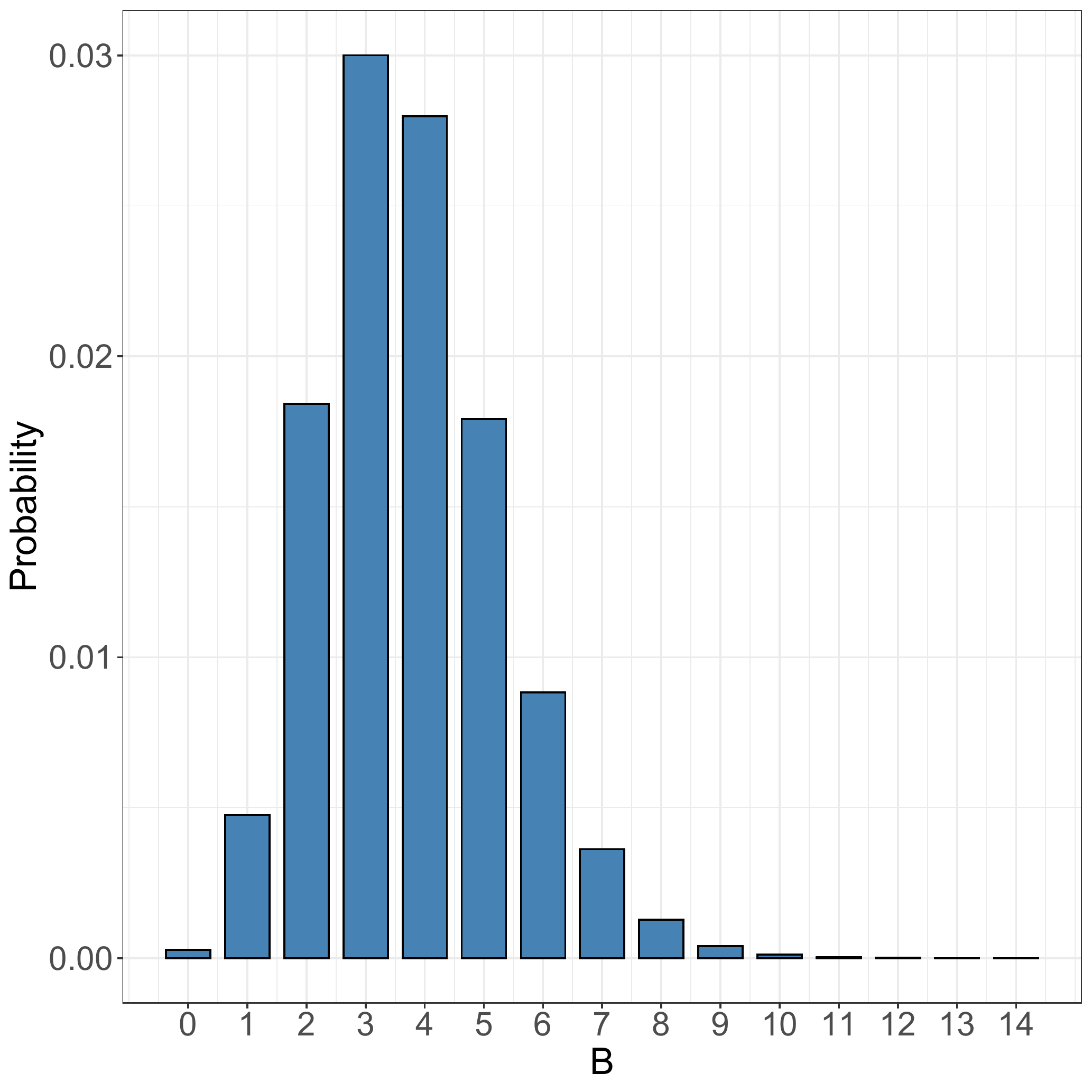}
	}
	\subfigure[]{
		\includegraphics[width=0.31\textwidth,height=4.5cm]{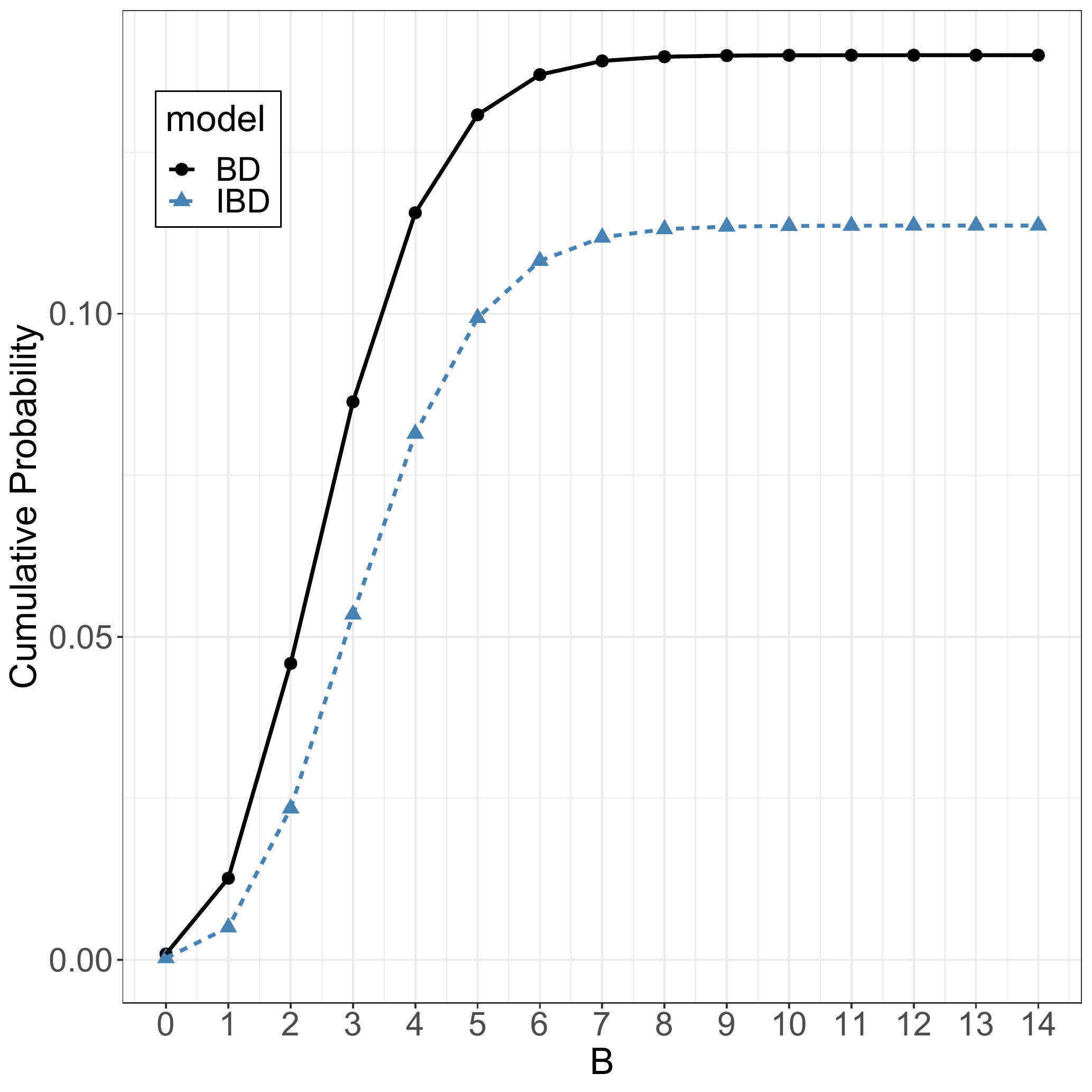}
	}
	\caption{ (a) Estimates for $p_{5,5}^B(1)$ of the L-BDI process $(\nu=0)$. (b) Estimates for $p_{5,5}^B(1)$ of the L-BDI process $(\nu=1.2)$.   (c)  Cumulative probabilities.}
	\label{BD_range}
\end{figure}

\vspace{0.3cm}
The IGBS method could also be applied to estimate $p_{ij}^B(t)$. As an illustration,  $p_{5,5}^B(1)$  of the L-BDI process with immigration rate $\nu =0$ and $1.2$ respectively are evaluated $(B=0,\ldots,14)$. The sample size is $10^5$ and the results are plotted in Fig. \ref{BD_range}.  

\begin{figure}[!ht]
	\centering
	\subfigure[]{
		\includegraphics[width=0.4\textwidth,height=3.5cm]{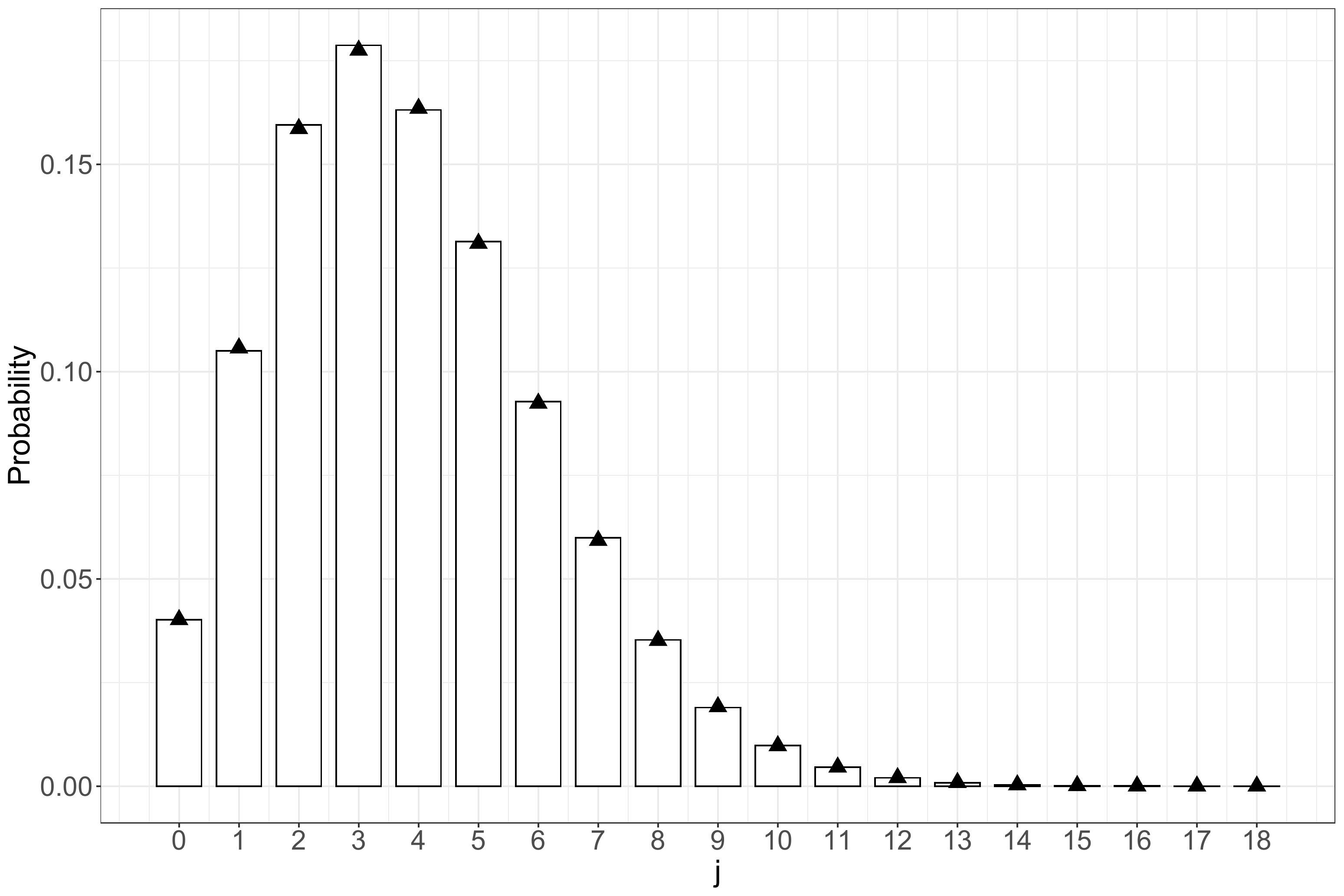}
	}
	\hfil
	\subfigure[]{
		\includegraphics[width=0.4\textwidth,height=3.5cm]{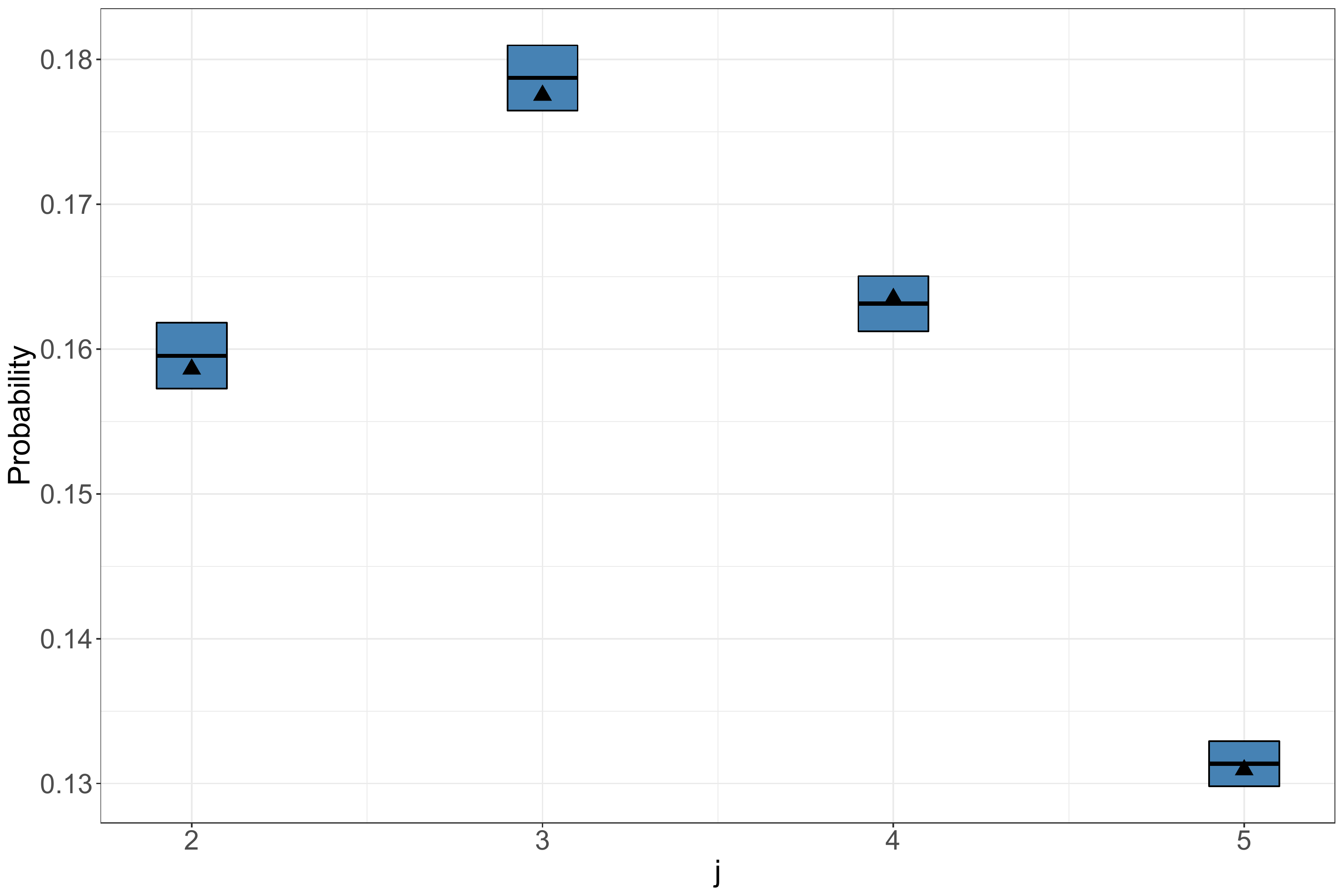}
	}
	\caption{Comparisons between the  estimates by the IGBS method and the straight simulation for the SIS model.  Crossbars  in (b) indicate the range of $\pm 2$ standard deviations.}
	\label{pij_sis}
\end{figure}

\newpage
For the SIS model, the IGBS method is applied to estimate its transition probabilities for $(t=1,i=5,j=0,\ldots,21)$. The results are displayed in Fig. \ref{pij_sis}. The black triangle represents the estimates given by simulating sample paths through the Gellispie method \citep{gillespie1977exact} and calculating the proportion of paths with terminal state $j$. This estimation procedure is referred as the straight simulation here. The sample size for the IGBS method is set as $10^5$ but $10^6$  for the straight simulation. The results are consistent as expected.

When the value of transition probability is   close to $0$, the time cost of the straight simulation to reach desired relative accuracy might go beyond tolerance. But the IGBS scheme is designed to sample the the bridge path, hence it is overwhelming
in the Monte Carlo evaluation of probabilities of rare events. As an numerical example,  the IGBS estimates are made for  epidemic termination probabilities in unit time of SIS model with different initial infectious individuals. The results are reported in Table \ref{sis_extinct}. The four significant digits value of the termination probability of the SIS model with $I_0=20$ is $8.948\times 10^{-6}$. To reach the same accuracy with the straight simulation, sample size up to $10^8$ is needed. That would be 100 times more than the IGBS scheme. In the case $I_0=30$, the time cost of straight simulation would be embarrassing.

\begin{table}[!ht]
	\caption{\it Estimates of the termination probability for the SIS model with sample size $10^6$.}
	\centering
	\begin{tabular}{rcc}
		\hline
		$I_0$ & $\hat{p}$ & $\hat{\sigma}$ \\ 
		\hline \\
		10 & $\qquad 1.997\times 10^{-3}$ & $\qquad 1.193\times 10^{-5}$ \\ 
		20 & $\qquad 8.948\times 10^{-6}$ & $\qquad 9.678\times 10^{-8}$  \\ 
		30 & $\qquad 8.529\times 10^{-8}$& $\qquad 1.151\times 10^{-9}$ \\ 
		\hline  \\
	\end{tabular}
	\label{sis_extinct}
\end{table}

\subsection{ The stochastic SIR model as a birth-death process}
The susceptible-infectious-removed, SIR, model originally proposed by \cite{kermack_contribution_1927} is another fundamental model in epidemic dynamics.
In the SIR model, individuals in a closed community are classified into three compartments, that is, the susceptibles ($S_t$), the infectious ($I_t$) and the removed ($R_t$). Besides, at each time $t$ we have $S_t+I_t+R_t\equiv N_0$, where $N_0$ is the number of individuals in the community. This identity implies that the entire system are governed by dynamical changes of $S_t$ and $I_t$. The SIR model assumes that individuals are well mixed in the community and a patient may come across every susceptible individual with the same probability in a sufficiently small time interval. Consequently, the increment in the infectious individuals in unit time will be proportional to $S_t I_t$, with rate parameter $\beta>0$. Meanwhile, with $1/\gamma$ as the average infected time, the reductions in the infectious individuals in unit time should be $\gamma I_t$. Hence, the SIR model in terms of ordinary differential equations follows: 
\begin{equation*}
	\dot{S_t} = -\beta S_t I_t,
	\qquad \dot{I_t} = \beta S_t I_t -\gamma I_t.\\
\end{equation*}
The model is merely a primary approximation to a real pandemics with large
infectious population where the integer counting and the random interaction are of little 
significance. \cite{bartlett1949some} generalized the SIR model to the form of birth-death processes. Different modern versions of the SIR modelling could be found in  \citep{britton_stochastic_2019}. Hereafter the SIR model is referred to its birth-death version and  the vector $Z=(S,I)$, where $S=\{S_t,t\geq 0\}$ and $I=\{I_t,t\geq 0\}$, denotes the process. 

The events and corresponding rate functions of the SIR model are given in Table \ref{table:sir}. Essentially, this model is a two-dimensional random walk.
The state space of the SIR model consists of the integer grid points in  the triangular region enclosed by the $I$-axis, $S$-axis and the straight line $S_t+I_t=N_0$ of which
the states on $S$-axis are absorbing. 
\begin{table}[!ht]
	\caption{\it The SIR model of the birth-death type.} 
	\centering
	\begin{tabular}{ccl}
		Type number & Event & $\qquad $ Rate function\\  \\
		1&$\qquad (S,I)\to(S-1,I+1)$ & $\qquad \lambda_1(S_t,I_t)=\beta S_t I_t$ \\
		2&$(S,I)\to(S,I-1)$ & $\qquad \lambda_2(S_t,I_t)=\gamma I_t $ \\
	\end{tabular}
	\label{table:sir}
\end{table}

With complete observations, the likelihood function of the SIR model can be 
written out explicitly as follows. Let $\omega=(\omega_S,\omega_I)$ be a two-dimensional path of $(S,I)$ connecting the states $Z_0=(S_0,I_0)$,
say $I_0=i$ and $I_t=j$, and $Z_t=(S_t,I_t)$ over time interval $[0,t]$. Assume
that $K$ events happened during this period.  For the cases with $S_0=0$ and $I_0>0$, the SIR model reduces to a pure death process of $I$. If $I_0=0$, the epidemic is terminated and  no event will happen subsequently. Excluding these two kinds of trivial cases, the initial states $S_0$ and $I_0$ are both assumed to be positive.
Further decompose the sample paths as $\omega_S=(\tau_{0:(K+1)},\omega_{S,0:(K+1)})$ and $\omega_I=(\tau_{0:(K+1)},\omega_{I,0:(K+1)})$. And let $x_{j,k}={\bf I}(\text{type $j$ transition in $\omega$ for $k$th event })$ for $j=1,2$. Then the likelihood function of $\omega$ can be written as 
\begin{equation}
	\label{eq:lki_sir}
	\begin{aligned}
		p(t,Z_0\overset{\omega}{\rightarrow}Z_t)=&\prod_{k=1}^{K}\left(\lambda_1(\omega_{S,k-1},\omega_{I,k-1})^{x_{1,k}} \lambda_2(\omega_{S,k-1},\omega_{I,k-1})^{x_{2,k}}\right)\cdot\\
		&\exp\left\{-\int_{0}^{t} (\lambda_1(\omega_{S,s},\omega_{I,s})+\lambda_2(\omega_{S,s},\omega_{I,s}))ds\right\}
	\end{aligned}
\end{equation}

Expression (\ref{eq:lki_sir}) suggests the estimation of $p(t,Z_0, Z_t)$ by the IGBS method. Noting that an upward jump in $\omega_I$ corresponds to the event of form $(S,I)\to(S-1,I+1)$ while a downward jump associates with $(S,I)\to(S,I-1)$,  $\omega_S$ can be reconstructed completely from the occurring times of upward jumps in $\omega_I$. Therefore 
the SIR model is equivalent to a birth-death process in terms of $I$.  Given $Z_0$ and $Z_t$, the number of upward jumps in $\omega_I$ and the transitions in $\omega_S$ is $B=S_0-S_t$ and the number of downward jumps $D=I_t-I_0+B=(S_0-I_0)-(S_t-I_t)$.  Denote  by $\Omega_{ij}^B(t)$ the set of $\omega_I$ with initial state $i$ and terminal state $j$ 
over time interval $[0,t]$. 

Now the Algorithm \ref{alg:IGBS} could be employed for the 
likelihood inference of the SIR model.

\subsection{An IGBS filter for incomplete birth-death records}
Often in practices, only partial records of $S$ and $I$ are available. Even worse, situations with only $S$ records or only $I$ records are by no means rare.  Statistical inference manipulating this kind of missing data model is of 
serious concern over the last few decades. 
The general principle is to establish certain  hidden Markov dynamics governing the evolution of the
unobserved  state process. Then efforts are
contributed to reconstructing the state process
with a posterior sampler. 
Particle filter is the most popular scheme among such posterior samplers.

To demonstrate the potential power of the IGBS method, this subsection is contributed to the construction  and application of a hybrid
algorithm  to perform the Bayesian inference for the SIR model when only
consecutive $S$ records at discrete time epochs are available. The
unobserved background birth-death state process now appears more challenging
as compared with the regular SIR 
model depicted in the previous subsection.  Since the generalized IGBS algorithm offers an alternative
for the particle filter, the title IGBS filter is tagged. 

Consider a set of $N+1$ observations of $S$, denoted by $S_{0:N}=(S_0,S_1,\ldots,S_N)$ recorded at instants $t_{0:N}=(t_0,t_1,\ldots,t_N)$. Let $I_k$ be  the unobserved values taken by $I$ at $t_k$, $k=0,1,\ldots,N$. Define $\Delta S_k = S_{k-1}-S_k$ to represent the increment in the infected individuals between the $k$th and the $(k+1)$th observations. Then $\{S_0,\, \Delta S_{1:N}=(\Delta S_1, \ldots,\Delta S_N)\}$ contains the same information as  $S_{0:N}$.

Given $(S_0,I_0)$ (usually $I_0=1$), the  likelihood function $L(\theta\mid S_{0:N})$ takes the form:
\begin{equation}
	\label{sir_p_de}
	L(\theta\mid S_{0:N} )=  \prod_{k=1}^{N} \text{pr}(S_k\mid S_{0:k-1} ).
\end{equation}	
Here pr$(S_k\mid S_{0:k-1},\, \theta)$ is written as 
pr$(S_k\mid S_{0:k-1} )$ for convenience.  Expression (\ref{sir_p_de}) implies that the evaluation of  $L(\theta\mid S_{0:N} )$ can be achieved by calculating the conditional probabilities pr$(S_k\mid S_{0:k-1}),\; k=1,\ldots,N$ recursively. This is possible
due to the following proposition.  

\vspace{.3cm}
\begin{proposition} 
	\label{prop_IGBS_filter}
	The conditional probability $ \text{pr}(S_k\mid S_{0:k-1})$ can be
	evaluated as the mathematical expectation over bridge path spaces:
	\begin{equation}
		\label{seq3} 
		\text{pr}(S_k\mid S_{0:k-1}) =  p 
		E \left[
		q_{ij}^B(\omega)\mid S_{0:k},\, I_{k-1}>0 \right]  
		+(1-p)  {\bf  I}(S_k = S_{k-1}).
	\end{equation}
	\begin{eqnarray*}
		\text{ where } \hspace{1cm} p& = &P(I_{k-1}>0|S_{0:k-1}),  \\
		q_{ij}^B(\omega) & = &  p\left(\Delta t_k,i\overset{\omega}{\rightarrow}j \right)  \frac{B+i+1}{h_{ij}^B(\Delta t_k)}, \hspace{0.5cm}
		\underline{(i=I_{k-1},\, j=I_k,\, B = \Delta S_k)}.
	\end{eqnarray*}
\end{proposition}

\noindent {\bf Proof:}
\begin{equation}
	\label{sir0}
	\text{pr}(S_k\mid S_{0:k-1}) = \sum_{I_k} \text{pr}(S_k,I_k\mid S_{0:k-1}). 
\end{equation}
Now that the $(S,I)$ as a process is Markovian, 
\begin{equation}
	\label{sir1}
	\text{pr}(S_k,I_k\mid S_{0:k-1})  = \sum_{I_{k-1}} \text{pr}(S_k,I_k\mid S_{k-1},I_{k-1})\text{pr}(I_{k-1}\mid S_{0:k-1}). 
\end{equation}
Thus
\begin{equation}
	\label{sir2}
	\text{pr}(S_k\mid S_{0:k-1}) =\sum_{I_k,\,I_{k-1}} \text{pr}(S_k,I_k\mid S_{k-1},I_{k-1})\text{pr}(I_{k-1}\mid S_{0:k-1}).
\end{equation}
This is the hardcore  issue:	$\text{pr}(S_k,I_k\mid S_{k-1},I_{k-1})$ and $\text{pr}(I_{k-1}\mid S_{0:k-1})$ do not have explicit expressions in general. Even if they do in some special situations, the summation could 
hardly bring about a simple formula.   So usually (\ref{sir2}) cannot be evaluated directly . Hence the posterior sampling for $ (I_{k-1},
\, I_k)$ given $S_{0:k}$  is 
the most serious challenge of grave concern. Whereas, as shown in the 
following, the IGBS method with minor modification could be employed 
as a numerical algorithm for    (\ref{sir2}). 

Noting that,
\begin{eqnarray*}
	\text{pr}(S_k,I_k\mid S_{k-1},I_{k-1})& = &\int p(\Delta t_k, (S_{k-1},I_{k-1})\overset{\omega}{\rightarrow}(S_k, I_k))\;d\omega,\\
	& = &\int   p(\Delta t_k,i\overset{\omega}{\rightarrow}j )
	\text{\huge $\mid $}_{(i=I_{k-1},\, j=I_k)}^{B = \Delta S_k}\; d\omega, 
\end{eqnarray*}
in the manner of IGBS method, $\text{pr}(S_k\mid S_{0:k-1})$ can be rewritten as
\begin{eqnarray}
	\label{seq}
	\text{pr}(S_k\mid S_{0:k-1}) & = &\text{\Large $\sum_{I_{k-1},\,I_k}$}
	\text{ \LARGE $\int$ }  \underline{ \left[
		\frac{ p\left(\Delta t_k,i\overset{\omega}{\rightarrow}j \right)}
		{h^B_{ij}(\Delta t_k)}\;\cdot h^B_{ij}(\Delta t_k) \cdot \text{pr}(I_{k-1}\mid S_{0:k-1})\right]}\;  d\omega,  \nonumber   \\
	& & \hspace{3cm}(i=I_{k-1},\, j=I_k,\, B = \Delta S_k) 
\end{eqnarray}
where $ h^B_{ij}(\Delta t_k)$ is given by the formula previously in this paper. 
If the $t_k$-prior distribution $\text{pr}(I_{k-1}\mid S_{0:k-1})$ is available, i.e. samples of $I_{k-1}$ can be readily generated,  then  expression (\ref{seq}) is almost ready for use but for lacking a sampler of $I_k$. Given $\{I_{k-1}=i$, $S_{k-1}$ and $S_k\}$,  maximum possible value of $I_k$ should be 
$I_{k-1}+ \Delta S_k= i + B$. Thus the set of all possible values of
$I_k=j$ is
\begin{equation*}
	{\cal A}_k = \{0,1,\ldots,I_{k-1}+\Delta S_k\}= \{0,1,\ldots,i+B\}.
\end{equation*}
Now the uniform distribution over ${\cal A}_k$ would facilitate   the whole scheme a clean take-off. 

In addition, it is more efficient to handle the cases of $I_{k-1}=0$ and $I_{k-1}> 0$ separately. Because once $I$ hits zero, the epidemic is terminated and $S$ would not vary any more. So if  $I_{k-1}=0$, the transitional probability would take the 0-1 form
\begin{equation*}
	\text{pr}(S_k,I_k\mid S_{k-1},I_{k-1}=0)=\begin{cases}
		1,\;\;\;\;I_k=0\,\text{ and }\, S_k=S_{k-1}; \\
		0,\;\;\;\;\text{otherwise}.
	\end{cases}
\end{equation*}
Therefore samples of $I_{k-1}=i$ should be drawn from  
$\text{pr}(I_{k-1}\mid S_{0:k-1},I_{k-1}>0)$
instead of $\text{pr}(I_{k-1}\mid S_{0:k-1})$, and expression (\ref{seq}) could be reformulated as 
\begin{equation}
	\label{seq2}
	\begin{aligned}
		\text{pr}(S_k\mid S_{0:k-1})=& \text{pr}(I_{k-1}>0\mid S_{0:k-1}) \cdot 
		\text{\large $\sum_{I_{k-1},\,I_k} $}\left[
		\int  \underline{q_{ij}^B(\omega)\frac{h_{ij}^B(\Delta t_k)}{ B+i+1}\text{pr}(I_{k-1}\mid S_{0:k-1},I_{k-1}>0)} d\omega \right] \\
		&+\text{pr}(I_{k-1}=0\mid S_{0:k-1}) \cdot \text{pr}(S_k,I_k\mid S_{k-1},I_{k-1}=0)
		\cdot {\bf  I}(S_k = S_{k-1}) ,
	\end{aligned}
\end{equation}
\begin{equation*}
	\text{where}\hspace{1cm}	q_{ij}^B(\omega) =  p\left(\Delta t_k,i\overset{\omega}{\rightarrow}j \right)\frac{B+i+1}{h_{ij}^B(\Delta t_k)}, \hspace{0.5cm}
	\underline{(i=I_{k-1},\, j=I_k,\, B = \Delta S_k)}.
\end{equation*}

With the sequential sampling in mind:
\begin{tabbing} 
	\hspace{2cm} \=$I_{k-1}$ \= $\sim$ \= $\text{pr}(I_{k-1}\mid S_{0:k-1},I_{k-1}>0)$,\; \=$ i = I_{k-1}$; \\ \\
	\>$I_k$ \> $ \sim  $ \> U$({\cal A}_k)$, \; $B=\Delta S_k$, \;
	\>$j = I_k$,\; $l=0$,\; $u=B+i+1$;\\ \\
	\>$ \omega $ \> $\sim  $ \>U$(\Omega_{ij}^B(\Delta t_k))$, 
	\>by IGBS;\\
\end{tabbing}
the expression (\ref{seq2}) takes a fairly compact form (\ref{seq3}):
\begin{equation*}
	\label{seq3} 
	\text{pr}(S_k\mid S_{0:k-1}) =  p 
	E \left[
	q_{ij}^B(\omega)\mid S_{0:k},\, I_{k-1}>0 \right]  
	+(1-p)  {\bf  I}(S_k = S_{k-1}).
\end{equation*}
\begin{eqnarray*}
	\text{ where } \hspace{1cm} p& = &\text{pr}(I_{k-1}>0\mid S_{0:k-1}),  \\
	q_{ij}^B(\omega) & = &  p\left(\Delta t_k,i\overset{\omega}{\rightarrow}j \right)  \frac{B+i+1}{h_{ij}^B(\Delta t_k)}, \hspace{0.5cm}
	\underline{(i=I_{k-1},\, j=I_k,\, B = \Delta S_k)}.  
\end{eqnarray*}
This completes the proof.
\vspace{.3cm} 

Proposition \ref{prop_IGBS_filter}  leads to the following Algorithm \ref{alg_IGBS_Filter}.

\begin{algorithm}
	\caption{ IGBS filter}
	\label{alg_IGBS_Filter} 
	\begin{tabbing}
		For $m=1:M$,\\
		\hspace{.5cm} \=Draw \hspace{.3cm} \=$I_{k-1}^{(m)}\sim 
		P(I_{k-1}\mid S_{0:k-1},\,I_{k-1}>0). $ \\
		\>$\;\;\;\;$ Set \>$\;\;\;\;  i\leftarrow I^{(m)}_{k-1}, \;\; 
		B\leftarrow S_{k-1}-S_k,\;\; ({\cal A}_k=\{0,1,\ldots, B+i\})$.\\
		\> Draw \>$I_k^{(m)}\sim$ U${(\cal A}_k)$.  \\
		\>$\;\;\;\;$ Set \>$\;\;\;\;  j\leftarrow I_k^{(m)},\;\;\;\;$
		$l\leftarrow 0,\;\; u\leftarrow B+i+1$; $\;\;$ 
		\\
		\>Draw \>$\omega^{(m)} \sim $ U$\left(\Omega_{ij}^B(\Delta  t_k)\right), \;\;$ by Algorithm \ref{alg:IGBS};\\
		\>Calculate and record $\;\; q^{(m)}\leftarrow q_{ij}^B(\omega^{(m)})$.\\
		End for $m$ 
	\end{tabbing}
	\vspace*{-10pt}
\end{algorithm}

\vspace{0.2cm}

The Mote Carlo estimator of pr$(S_k\mid S_{0:k-1})$ based upon Algorithm 
\ref{alg_IGBS_Filter} takes the form
\begin{equation}
	\begin{aligned}
		\text{pr}(S_k\mid S_{0:k-1})\approx \; & p 
		\left[\frac{1}{M}\sum_{m=1}^{M} q^{(m)}\right]  + (1-p)  
		{\bf  I}(S_k = S_{k-1}),\\
		& \text{ where }\;\;\;\; p=\text{pr}(I_{k-1}>0\mid S_{0:k-1}).
	\end{aligned}
\end{equation}
\vspace{.3cm}

As long as the local prior distribution pr$(I_{k-1}\mid S_{0:k-1})$ could be updated by pr$(I_k\mid S_{0:k})$, the recursive calling of Algorithm 
\ref{alg_IGBS_Filter} would bring out the required likelihood evaluation.
With large $M$ value, the empirical local posterior distribution would
satisfy this purpose:
\begin{equation}  
	\begin{aligned}
		\text{pr}(I_k=j\mid S_{0:k})  \approx \; &   p   \left[\frac{1}{M}\sum_{m=1}^{M} q^{(m)} {\bf I}(I_k^{(m)}=j) \right] 
		\;    + (1-p)  {\bf  I}(j=0),  \;\;\;\; j = 0,1,\ldots\\
		& \text{ where }\;\;\;\; p=\text{pr}(I_{k-1}>0\mid S_{0:k-1}).
	\end{aligned}
\end{equation}

\subsection{Application to the Shigellosis outbreak data set}
\begin{figure}[!ht]
	
	\subfigure[]{
		\includegraphics[width=0.6\textwidth,height=5cm]{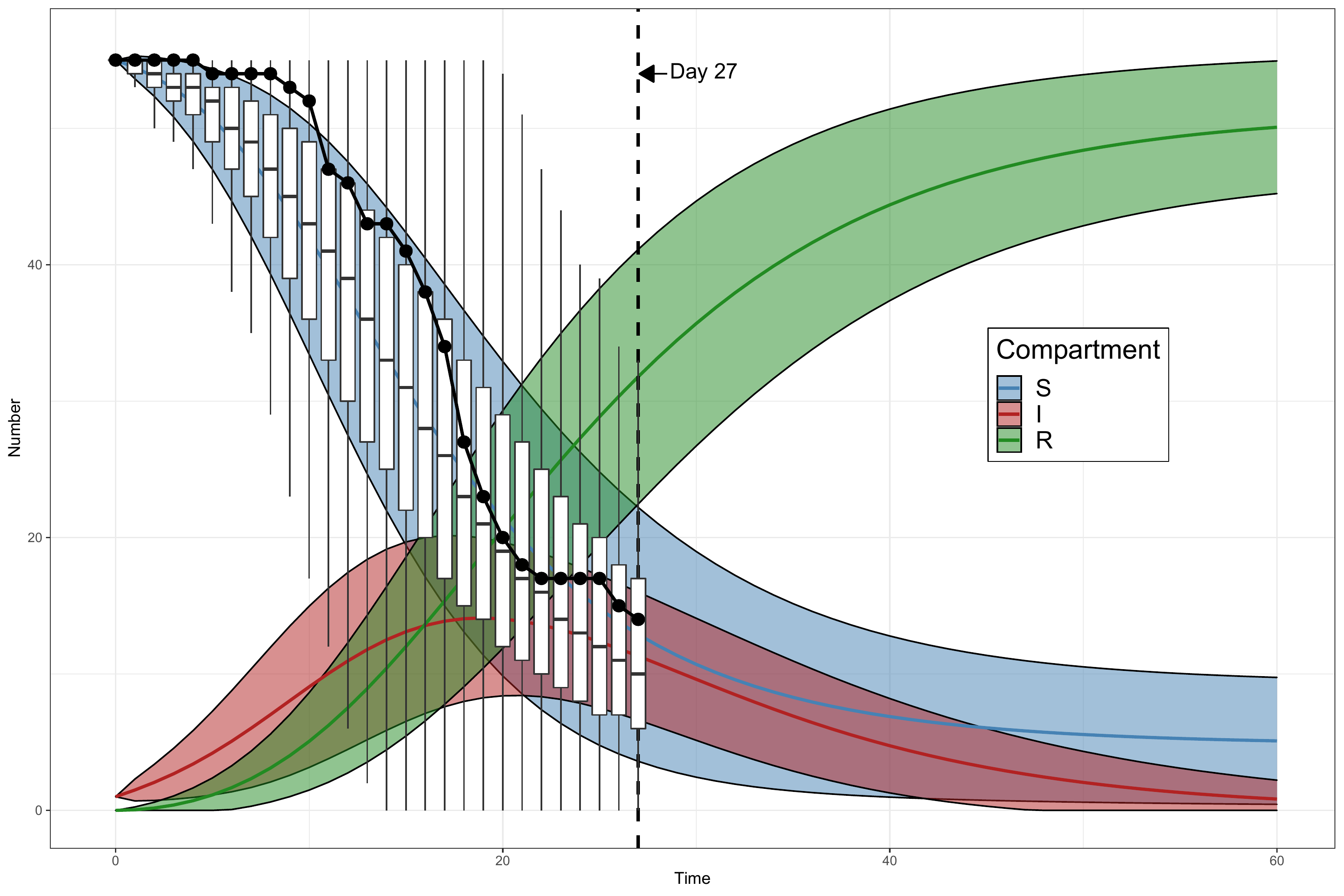}
	}
	\hfil
	\subfigure[]{
		\includegraphics[width=0.35\textwidth,height=6cm]{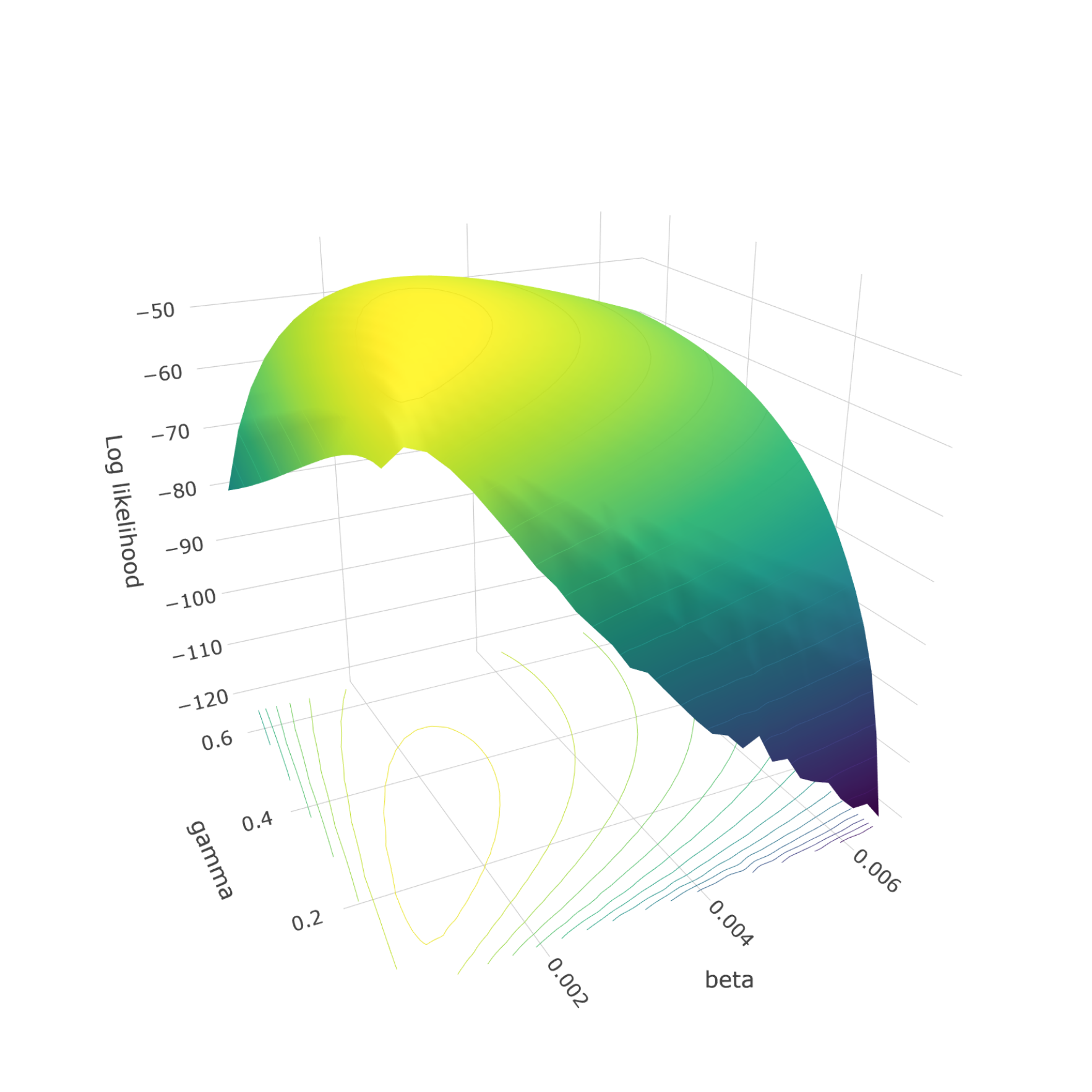}
	}\\
	
	\caption{Shigellosis analysis. (a) Reconstructed SIR
		paths. (b) The log-likelihood surface.
	}
	\label{fig:sir}
\end{figure}

Now   the IGBS filter  introduced above is applied to evaluate the maximum likelihood estimates of the parameters in SIR modelling for a real epidemic event data set, which records a Shigellosis outbreak in a shelter for the homeless in San Francisco from December 27, 1991 to January 23, 1992. The data comes from \cite{britton2002bayesian}.
The resulted SIR curves and the log-likelihood surface are plotted in Fig. \ref{fig:sir}. The  estimates of parameters  are reported in Table \ref{table:sir_mle}. 

\begin{table}[ht] 
	\centering
	\caption{\it Shigellosis Outbreak Data}
	\begin{tabular}{crrrrrrrrrrrrrr}
		\hline
		time & 0 & 1 & 2 & 3 & 4 & 5 & 6 & 7 & 8 & 9 & 10 & 11 & 12 & 13   \\ 
		S& 198 & 198 & 198 & 198 & 198 & 197 & 197 & 197 & 197 & 196 & 195 & 190 & 189 & 186\\
		\\
		time& 14& 15 & 16 & 17 & 18 & 19 & 20 & 21 & 22 & 23 & 24 & 25 & 26 & 27 \\ 
		S & 186 & 184 & 181 & 177 & 170 & 166 & 163 & 161 & 160 & 160 & 160 & 160 & 158 & 157 \\ 	 
		\hline
	\end{tabular} 
\end{table}

\begin{table}[!ht]
	\centering
	\caption{\it Maximum likelihood estimates for the SIR modelling of Shigellosis dataset}
	\begin{tabular}{ccc}
		\hline
		parameter&estimate &  $95\%$ CI \\  \\
		\hline
		$\beta$& 0.0016 & (0.0011, 0.0024) \\ 
		$\gamma$&0.2607 & (0.1624, 0.4032) \\ 
		\hline \\
	\end{tabular} 
	\label{table:sir_mle}
\end{table}

\vspace{0.2cm}
In \cite{britton2002bayesian}, this data set is modelled by a stochastic epidemic model with social structures. Here the data set is treated as a
severely incomplete stochastic SIR  samples.  
The posterior mean of the basic reproduction number $R_0$ reported by \cite{britton2002bayesian} is 1.12, while the maximum likelihood estimate of $R_0$ obtained by IGBS filter is 1.239, reasonably agreeable for an elementary modelling.

\subsection{Exorcise the filtering failure}

Birth-death processes with incomplete observations can be handled in the framework of state-space models, also known as the hidden Markov models. Such models consist of two layers, one is of the observational variables representing the measurements of the system, and the other layer consists of the hidden  state variables governing the evolution. In convention, the hidden state variable is denoted by $X$ and the observables by $Y$. Let $X_n$ and $Y_n$ be the true
values of $X$ and $Y$ at observation epoch $t_n$ respectively, then a state-space model could be expressed as:
\begin{equation*}
	X_0\sim \pi_0,\;\;\;\;\; \text{pr}(X_k\mid X_{k-1})= f(X_k\mid X_{k-1}),\;\;\;\;\; \text{pr}(Y_k\mid X_k)= g(Y_k\mid X_k),
\end{equation*}
where $\pi_0$ is the initial distribution of $X_0$, $f$ is the transition kernel of $X$ and $g$ is the conditional probability linking $X$ and $Y$. Besides, the filter distribution, i.e.  the distribution of $X_k$ conditioned upon $Y_{0:k}$, will be denoted by $\pi_k(X_k\mid Y_{0:k})$.

Particularly, in the context of SIR model for the Shigellosis data set, the hidden state variable is $X=I$. The observations  are daily reports of newly infected individuals. So $Y=S$ is taken.  This is a more general hidden Markov model with evolution pattern:
\begin{equation*}
	X_0\sim \pi_0,\;\;\;\;\; X_k|X_{k-1}\sim f(X_k|X_{k-1},Y_{k-1}),\;\;\;\;\; Y_k|X_k\sim g(Y_k|X_k).
\end{equation*}

Usually the posterior distribution  for  $(X_k\mid X_{0:k-1},\, Y_{0:k})$  is difficult to evaluate, so the Monte
Carlo schemes are favoured for inference. The most 
popular technique is still the particle filter.

Algorithm \ref{alg:pf} describes how the basic   bootstrap filter \citep{1993Novel,fearnhead_particle_2018} works.

\begin{algorithm} 
	\caption{Basic bootstrap filter (a sequential importance resampling filter)}
	\label{alg:pf}
	\begin{tabbing}
		\enspace For k = 1 to N\\
		\qquad For m = 1 to M\\
		\qquad\quad Draw $X_{k-1}^{(m)}\sim \hat\pi_{k-1}(X_{k-1}\mid Y_{0:k-1})$. \\
		\qquad\quad Draw $X_k^{(m)}\sim f(X_k\mid X_{k-1}^{(m)})$.\\
		\qquad\quad Calculate weight $w_k^{(m)} \leftarrow g(Y_k\mid X_k^{(m)})$.\\  
		\qquad End for $m$\\
		\qquad \quad Estimate conditional likelihood $\hat{P}(Y_k\mid Y_{0:k-1}) \leftarrow \sum_{m=1}^M w_k^{(m)}/M$.\\ 
		\qquad 	\quad Set $\hat{\pi}_{k}(z \mid Y_{0:k}) \leftarrow \;  
		\left(\sum_{m=1}^M w_k^{(m)}{\bf I}(X_k^{(m)}=z)\right)/\sum_{m=1}^M w_k^{(m)},\;\;$  
		for all proper $z$.\\
		\enspace End for $k$ \\
		\enspace Calculate the likelihood $L(\theta\mid Y_{0:N})\leftarrow \prod_{k=1}^N\hat{P}(Y_k\mid Y_{0:k-1})$.
	\end{tabbing} 
	\vspace*{-10pt}
	\vspace{.03cm}
\end{algorithm}
\vspace{.3cm}

However, there is no guarantee that Algorithm \ref{alg:pf} would work 
properly in general.  
Filtering failure \citep{stocks_iterated_2018} characterized by the singular posterior
measure is a common issue.

\begin{figure}[!htpb]
	\centering
	\subfigure[]{
		\includegraphics[width=0.4\textwidth,height=5cm]{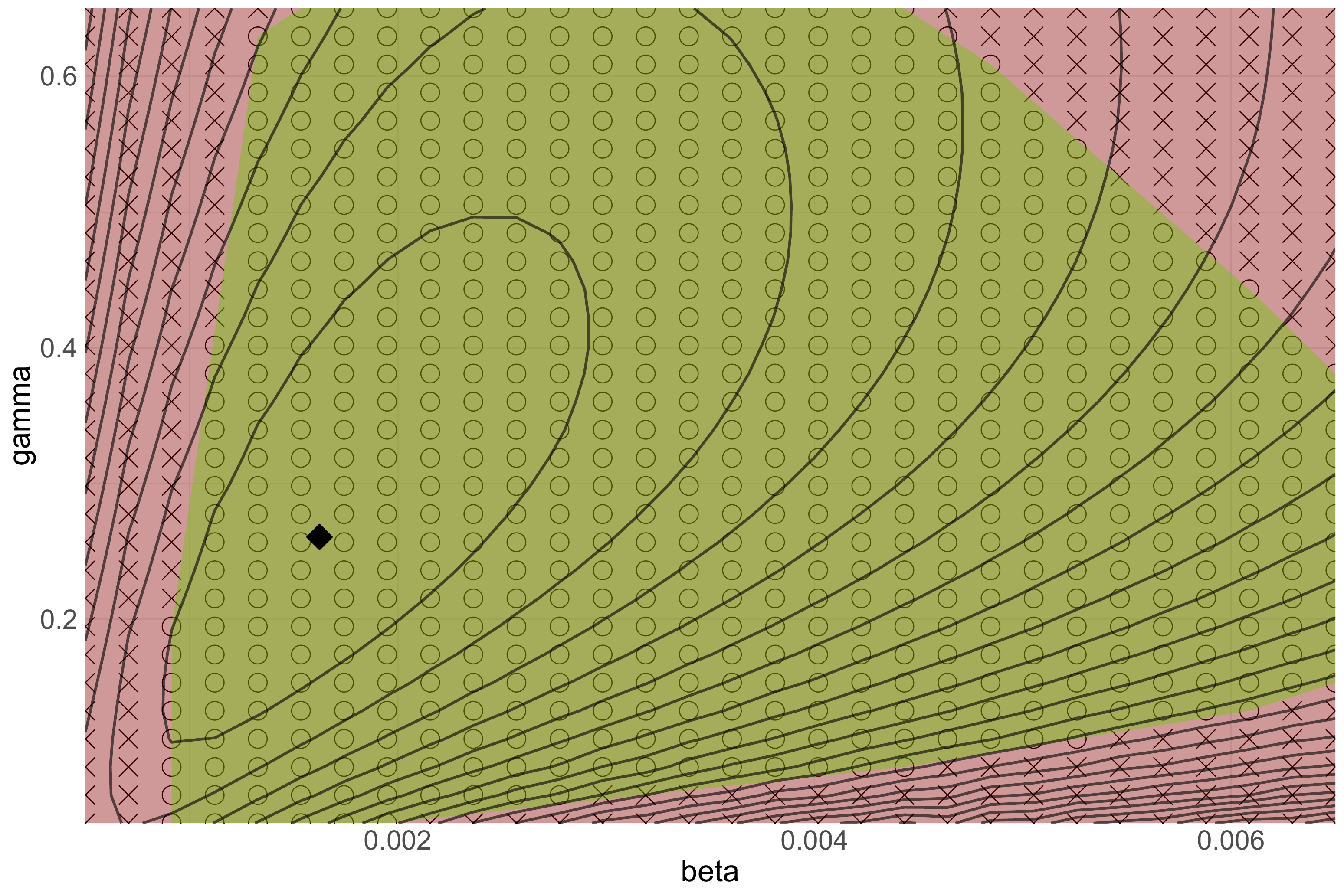}
	}
	\hfil
	\subfigure[]{
		\includegraphics[width=0.4\textwidth,height=5cm]{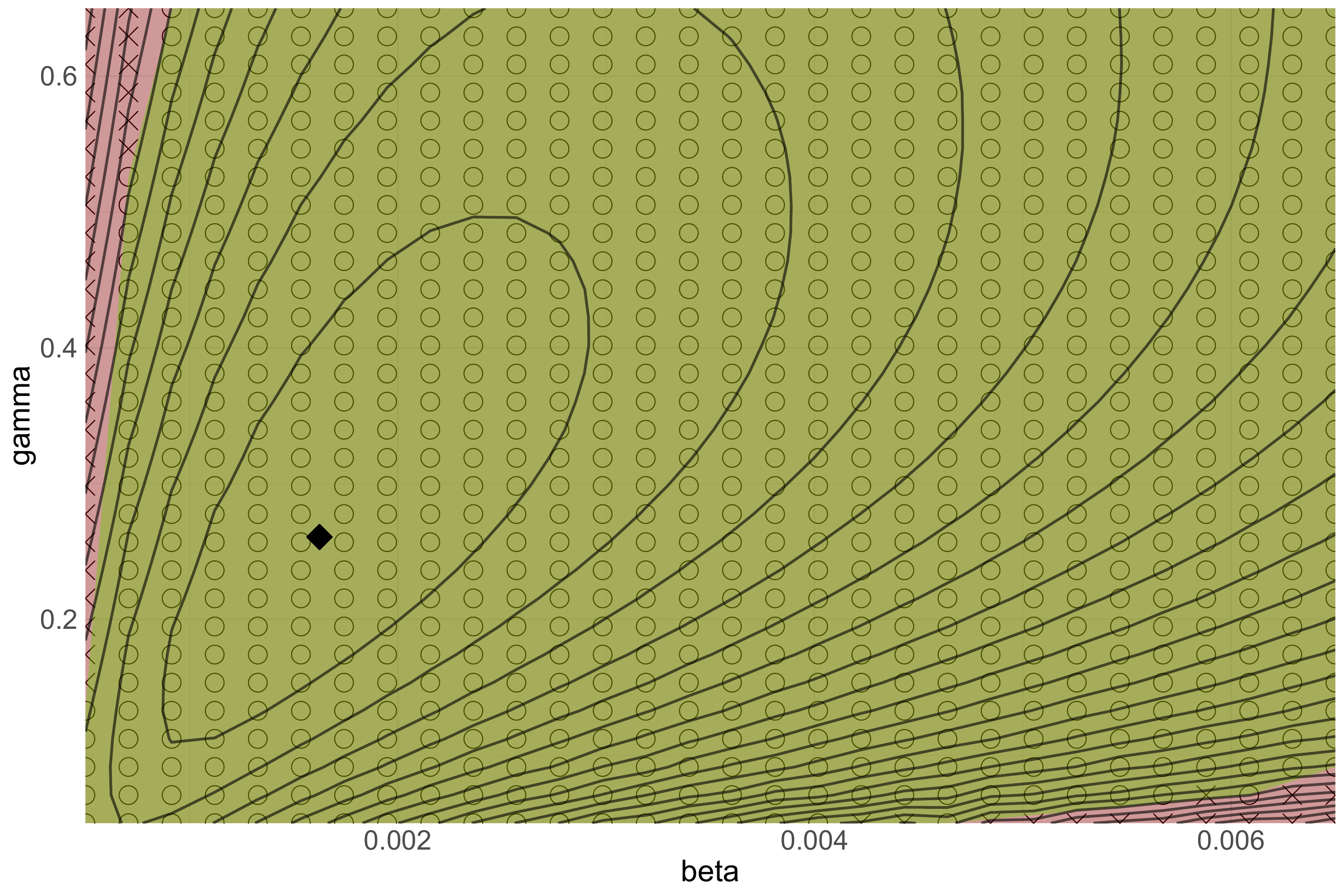}
	}
	\caption{Filter collapses encountered in SIR modelling of Shigellosis data. Parameters in pink regions would induce filtering failures. (a) Survival rate $<0.1 \% $. (b) Survival rate $<0.01 \% $.}
	\label{Fig:hit_mat}
\end{figure}

As an illustration for the filtering failure problem, Algorithm \ref{alg:pf} was also applied to the SIR modelling of Shigellosis data set. The particle number
is taken as one  million. During the performance, if the total number of particles with non-zero weights  falls below threshold, 
the corresponding parameters will be marked as a point in failure domain.   Figure \ref{Fig:hit_mat} depicts the failure
domains for thresholds $0.1\%$ and $0.01\%$ respectively.

The filtering failure usually happens in marginal areas, agreeing with the speculation that the causal fact is the
small likelihood values. Consequently, the maximum likelihood estimates might not be realized
properly if the initial values of parameters fall in the failure domains.

But for the IGBS scheme advocated here, sample paths are drawn from the bridge path space with uniform distributions free of the model parameters.   Robust evaluation of the likelihood could be obtained even in the failure domains of particle filters.

\section{Conclusion}
The integer grid bridge sampler, IGBS, proposed in the present paper was
constructed as a handy technique for the Bayesian inference
of general birth-death processes. The innovative idea is to establish
a one-to-one correspondence between the restricted birth-death bridge path space and 
the product space of the
integer grid bridge path set and the temporal simplex. Then the   sampling in the latter regular space naturally leads to a Monte Carlo scheme for the posterior 
sampling over the incomplete birth-death records. 

The effectiveness of the IGBS is shown with a few popular  models and in
the SIR modelling of the data set attributing to a real epidemic event. 
In principle  the IGBS method is applicable to general multi-dimensional birth-death
processes, say predator-prey model, without serious technical curse.

Fatal traps haunting the popular schemes like particle filters would
not hinder the IGBS simply because of the new sampler's non-parametric feature and essential ergodicity. The mismatched proposal distribution for the importance sampler  
embedded in particle filters invites the danger of measure distortion with each prior-posterior updating. Such filtering failures find no 
counterparts in the IGBS filter in general settings.

\section*{Acknowledgement}
Gratitude is owned to Dr. Peter Clifford, Oxford, U.K., for his deep insights and
encouragement over time and space.

\bibliographystyle{biometrika}
\bibliography{BD_ref}

\begin{thebibliography}{23}
\expandafter\ifx\csname natexlab\endcsname\relax\def\natexlab#1{#1}\fi

\bibitem[{Allen(2010)}]{allen2010introduction}
\textsc{Allen, L.~J.} (2010).
\newblock \textit{An introduction to stochastic processes with applications to
  biology}.
\newblock Chapman and Hall/CRC.

\bibitem[{Bailey(1990)}]{bailey1990elements}
\textsc{Bailey, N.~T.} (1990).
\newblock \textit{The elements of stochastic processes with applications to the
  natural sciences}, vol.~25.
\newblock John Wiley \& Sons.

\bibitem[{Bartlett(1949)}]{bartlett1949some}
\textsc{Bartlett, M.} (1949).
\newblock Some evolutionary stochastic processes.
\newblock \textit{Journal of the Royal Statistical Society: Series B
  (Methodological)} \textbf{11}, 211--229.

\bibitem[{Britton \& O'NEILL(2002)}]{britton2002bayesian}
\textsc{Britton, T.} \& \textsc{O'NEILL, P.~D.} (2002).
\newblock Bayesian inference for stochastic epidemics in populations with
  random social structure.
\newblock \textit{Scandinavian Journal of Statistics} \textbf{29}, 375--390.

\bibitem[{Britton \& Pardoux(2019)}]{britton_stochastic_2019}
\textsc{Britton, T.} \& \textsc{Pardoux, E.} (2019).
\newblock Stochastic epidemics in a homogeneous community.
\newblock \textit{{arXiv}:1808.05350 [math]} \textbf{2255}.

\bibitem[{Crawford et~al.(2018)Crawford, Ho \&
  Suchard}]{crawford_computational_2018}
\textsc{Crawford, F.~W.}, \textsc{Ho, L. S.~T.} \& \textsc{Suchard, M.~A.}
  (2018).
\newblock Computational methods for birth-death processes.
\newblock \textit{Wiley Interdisciplinary Reviews: Computational Statistics}
  \textbf{10}, e1423.

\bibitem[{Crawford et~al.(2014)Crawford, Minin \&
  Suchard}]{crawford_estimation_2014}
\textsc{Crawford, F.~W.}, \textsc{Minin, V.~N.} \& \textsc{Suchard, M.~A.}
  (2014).
\newblock Estimation for {General} {Birth}-{Death} {Processes}.
\newblock \textit{Journal of the American Statistical Association}
  \textbf{109}, 730--747.

\bibitem[{Crawford \& Suchard(2012)}]{crawford_transition_2012}
\textsc{Crawford, F.~W.} \& \textsc{Suchard, M.~A.} (2012).
\newblock Transition probabilities for general birth–death processes with
  applications in ecology, genetics, and evolution.
\newblock \textit{Journal of Mathematical Biology} \textbf{65}, 553--580.

\bibitem[{Doucet \& Johansen(2009)}]{doucet2009tutorial}
\textsc{Doucet, A.} \& \textsc{Johansen, A.~M.} (2009).
\newblock A tutorial on particle filtering and smoothing: Fifteen years later.
\newblock \textit{Handbook of nonlinear filtering} \textbf{12}, 3.

\bibitem[{Fearnhead \& Künsch(2018)}]{fearnhead_particle_2018}
\textsc{Fearnhead, P.} \& \textsc{Künsch, H.~R.} (2018).
\newblock Particle filters and data assimilation.
\newblock \textit{Annual Review of Statistics and Its Application} \textbf{5},
  421--449.

\bibitem[{Gillespie(1977)}]{gillespie1977exact}
\textsc{Gillespie, D.~T.} (1977).
\newblock Exact stochastic simulation of coupled chemical reactions.
\newblock \textit{The journal of physical chemistry} \textbf{81}, 2340--2361.

\bibitem[{Gordon \& Salmond(1993)}]{1993Novel}
\textsc{Gordon, N.~J.} \& \textsc{Salmond, D.~J.} (1993).
\newblock Novel approach to nonlinear/non-gaussian bayesian state estimation.
\newblock \textit{IEE Proceedings. Part F} \textbf{140}, P.107--113.

\bibitem[{Ho et~al.(2018{\natexlab{a}})Ho, Crawford \&
  Suchard}]{ho_direct_2018}
\textsc{Ho, L. S.~T.}, \textsc{Crawford, F.~W.} \& \textsc{Suchard, M.~A.}
  (2018{\natexlab{a}}).
\newblock Direct likelihood-based inference for discretely observed stochastic
  compartmental models of infectious disease.
\newblock \textit{The Annals of Applied Statistics} \textbf{12}, 1993--2021.

\bibitem[{Ho et~al.(2018{\natexlab{b}})Ho, Xu, Crawford, Minin \&
  Suchard}]{ho_birth/birth-death_2018}
\textsc{Ho, L. S.~T.}, \textsc{Xu, J.}, \textsc{Crawford, F.~W.},
  \textsc{Minin, V.~N.} \& \textsc{Suchard, M.~A.} (2018{\natexlab{b}}).
\newblock Birth/birth-death processes and their computable transition
  probabilities with biological applications.
\newblock \textit{Journal of Mathematical Biology} \textbf{76}, 911--944.

\bibitem[{Kermack \& McKendrick(1927)}]{kermack_contribution_1927}
\textsc{Kermack, W.~O.} \& \textsc{McKendrick, A.~G.} (1927).
\newblock A contribution to the mathematical theory of epidemics.
\newblock \textit{Proceedings of the Royal Society of London. Series A,
  Containing Papers of a Mathematical and Physical Character} \textbf{115},
  700--721.

\bibitem[{Kitagawa(1987)}]{kitagawa_non-gaussian_1987}
\textsc{Kitagawa, G.} (1987).
\newblock Non-{Gaussian} {State} {Space} {Modeling} of {Nonstationary} {Time}
  {Series}.
\newblock \textit{Journal of the American Statistical Association} \textbf{82},
  1032--1041.

\bibitem[{Norris(1998)}]{norris_markov_1998}
\textsc{Norris, J.~R.} (1998).
\newblock \textit{Markov chains}.
\newblock Cambridge series on statistical and probabilistic mathematics.
  Cambridge, UK ; New York: Cambridge University Press, 1st ed.

\bibitem[{Novozhilov et~al.(2006)Novozhilov, Karev \&
  Koonin}]{novozhilov_biological_2006}
\textsc{Novozhilov, A.~S.}, \textsc{Karev, G.~P.} \& \textsc{Koonin, E.~V.}
  (2006).
\newblock Biological applications of the theory of birth-and-death processes.
\newblock \textit{Briefings in Bioinformatics} \textbf{7}, 70--85.

\bibitem[{Pfeuffer et~al.(2019)Pfeuffer, Möstel \&
  Fischer}]{pfeuffer_extended_2019}
\textsc{Pfeuffer, M.}, \textsc{Möstel, L.} \& \textsc{Fischer, M.} (2019).
\newblock An extended likelihood framework for modelling discretely observed
  credit rating transitions.
\newblock \textit{Quantitative Finance} \textbf{19}, 93--104.

\bibitem[{Renault(2008)}]{renault2008lost}
\textsc{Renault, M.} (2008).
\newblock Lost (and found) in translation: Andr{\'e}'s actual method and its
  application to the generalized ballot problem.
\newblock \textit{The American Mathematical Monthly} \textbf{115}, 358--363.

\bibitem[{Reynolds(1973)}]{reynolds_estimating_1973}
\textsc{Reynolds, J.~F.} (1973).
\newblock {ON} {ESTIMATING} {THE} {PARAMETERS} {OF} {A} {BIRTH}-{DEATH}
  {PROCESS}.
\newblock \textit{Australian Journal of Statistics} \textbf{15}, 35--43.

\bibitem[{Stocks(2018)}]{stocks_iterated_2018}
\textsc{Stocks, T.} (2018).
\newblock Iterated filtering methods for {Markov} process epidemic models.
\newblock \textit{arXiv:1712.03058 [math, stat]} ArXiv: 1712.03058.

\bibitem[{Stocks et~al.(2020)Stocks, Britton \& Höhle}]{stocks_model_2020}
\textsc{Stocks, T.}, \textsc{Britton, T.} \& \textsc{Höhle, M.} (2020).
\newblock Model selection and parameter estimation for dynamic epidemic models
  via iterated filtering: application to rotavirus in germany.
\newblock \textit{Biostatistics} \textbf{21}, 400--416.

\end{thebibliography}

\end{document}